\documentclass[%
 reprint,
 amsmath,amssymb,
 aps,twocolumn,
]{revtex4-2}
\usepackage{lineno}
\usepackage{xcolor}
\usepackage{longtable}
\usepackage{graphicx}% Include figure files
\usepackage{dcolumn}% Align table columns on decimal point
\usepackage{bm}% bold math
\usepackage{float}
\usepackage[sort&compress]{natbib}
\begin{document}
\title{Features of strangeness production in pp and heavy ion collisions}
\author{A. Pop}
\email{a$_$pop@nipne.ro}
\author{M. Petrovici} 
\affiliation{Horia Hulubei National Institute for Physics and Nuclear Engineering(IFIN-HH)\\
 Hadron Physics Department\\
RO-077125 Bucharest-Magurele, Romania}
\date{\today}

\begin{abstract}
Based on the existing experimental data for A-A collisions starting from the Alternating Gradient Synchrotron energies up to the CERN Large Hadron Collider  ones, various systematics related to strange hadrons and anti-hadrons are presented. As in the case of pions, kaons and protons, the ratio between the average transverse momentum and the square root
of the total particle multiplicity per unit rapidity and unit transverse overlap area $\langle p_{T} \rangle/\sqrt{\langle dN/dy \rangle/S_{\perp}}$ decreases with collision energy for a given centrality or with centrality for a given collision
energy, thus supporting the predictions of color glass condensate and percolation based approaches. The dependence on $\sqrt{\langle dN/dy \rangle/S_{\perp}}$ of the slope and offset extracted from the $\langle p_{T} \rangle$ - particle mass correlation and of the average transverse expansion velocity and kinetic freeze-out temperature parameters obtained from Boltzmann-Gibbs Blast Wave fits of the $p_{T}$ spectra for strange hadrons is compared to that for pions, kaons and protons, previously studied.

The detailed study of the entropy density ($\langle dN/dy \rangle/S_{\perp}$) dependence of the ratio of strange hadron yields per unit rapidity to the total particle multiplicity per unit rapidity ($Y^{S}/\langle dN/dy \rangle$) at different collision energies and centralities reveals the necessity to study
separately  strange hadrons and anti-hadrons. The correlation between  the ratio of the single- and multi- strange anti-hadron yield per unit rapidity to the total particle multiplicity per unit rapidity  and the  entropy density   is presented as a function of the fireball size.  A maximum is evidenced in the $Y^{S}/\langle dN/dy \rangle$ - $\langle dN/dy \rangle/S_{\perp}$ correlation  for combined and separate species of strange hadrons, at different centralities,  in the region where a transition from the baryon-dominated matter to the meson-dominated one takes place. Within the experimental error bars, the position of this maximum does not depend on the mass of the corresponding  strange hadron. Comparison with pp experimental data reveals another similarity between pp and Pb-Pb collisions at the CERN Large Hadron Collider  energies.
\end{abstract}

\maketitle

%\tableofcontents
\maketitle

\section{Introduction}
The possibility to produce hot and dense matter in heavy ion collisions \cite{Cha} such that, based on Quantum Chromo-Dynamics (QCD) asymptotic freedom properties, a transition from the hadronic phase to a high density "quark soup" \cite{Coll,Cab} or "quark-gluon plasma" \cite{Shu} can take place,  has motivated an unprecedented international effort in building accelerator facilities and complex experimental devices. The objects produced in such collisions have a size at the fermionic level, are highly inhomogeneous and undergo violent dynamics. Therefore, specific experimental probes have to be studied and theoretical approaches, combining different hypotheses for different stages of the formation and evolution of such systems produced in heavy ion collisions, are required for an unambiguous conclusion.
The first estimates of the transition from a gas of free nucleons to hadronic matter and subsequently to deconfined matter as a function of density were done within the percolation approach \cite{Bay,Cel}. Phenomenological models predicted some discontinuities in the behaviour of different observables as a function of collision energy or centrality specific for a phase transition between two thermodynamic states in a closed volume \cite{Van,Bla1,Bla2}. Recently, based on the existing experimental results from the Alternating Gradient Synchrotron (AGS), Super Proton Synchrotron (SPS), BNL Relativistic Heavy Ion Collider (RHIC) and CERN Large Hadron Collider (LHC), it was evidenced such a trend in the dependence of the ratio of the energy density to the entropy density ($\langle dE_{T}/dy \rangle/\langle dN/dy \rangle$) as a function of entropy density at different collision centralities for A-A collisions \cite{petro1}. The ratio of the energy density to the entropy density, at a given value of the transverse overlap area, increases with entropy. A tendency towards saturation at values of the entropy density beyond 6-8 $fm^{-2}$, corresponding to the largest collision energies at RHIC, and a steep rise at the LHC energies is evidenced for central collisions. Worth mentioning that for central collisions, a change in the collision energy dependence of the ratio $(1-R_{AA}^{\pi^{0}})/\langle dN/dy \rangle$ \cite{petro2} takes place in the same energy range as the one corresponding to the transition from an increase to the saturation in the entropy density dependence of $\langle dE_{T}/dy \rangle/\langle dN_{ch}/dy \rangle$. Such trends are in qualitative agreement with theoretical model predictions \cite{Bla1, Bla2} and \cite{liao,burke,Shi}, respectively. 40 years ago, well before the experimental data became available, the enhancement of the strangeness production was advocated as sensitive probe for deconfinement \cite{rafelski1}. A series of experiments from SPS to LHC energies evidenced an enhancement of strange hadron production as a function of centrality relative to the one corresponding to the pp minimum bias collision at the same energy. The influence of the core-corona relative contribution on the centrality dependence of the strangeness production, average transverse momenta, elliptic flow or $p_{T}$ spectra in heavy-ion collisions at SPS, RHIC and LHC energies was reported in many papers \cite{becattini1,bozek1,werner1,becattini2,becattini3,aich1,aich2,aich3,bozek2,stein,schreib,gemard1,Pet4}. As far as concerns the centrality dependence of the strange hadron production, it was clearly shown that this is the result of the interplay between the corona and core relative contributions. A consistent approach based on the string density before hadronization within the EPOS model using global parameters \cite{werner2} has shown that the core-corona contribution also explains  the similarity between Pb-Pb, p-Pb and pp in terms of strange hadron production as a function of charged particle density at mid-rapidity evidenced at LHC \cite{ALICEN}. A core-corona approach based on a new microcanonical hadronic recipe (EPOS4) \cite{wernerepos4} confirms the previous results. Once the experimental data  on strangeness production at LHC energies were available, it has been shown that the ratio of the strange hadron yield to entropy as a function of the entropy density, for the most central collisions \cite{petropop}, qualitatively follows a similar trend as the one observed in the  $\langle dE_{T}/dy \rangle/\langle dN/dy \rangle - \langle dN/dy \rangle/S_{\perp}$ correlation, indicative for a transition from a hadronic gas to a mixed phase of hadronic and deconfined matter. In the present paper, using the most complete experimental information from AGS to LHC energies, a detailed study of the ratio between the single- and multi- strange hadron and anti-hadron yields per unit rapidity and the total particle multiplicity per unit rapidity in relation to the entropy density, at different centralities and collision energies, is presented. Similarities between pp and Pb-Pb collisions at LHC energies are evidenced. Results on the entropy density dependence of observables related to the dynamics of the fireball are presented in Section II. Section III is dedicated to a detailed study of the entropy density dependence of the ratio between single- and multi- strange hadron and anti-hadron yields per unit rapidity and the total particle multiplicity per unit rapidity at different centralities and collision energies. In Section IV various aspects of such correlations are presented for different quark/antiquark compositions, fireball sizes and centralities. Similarities in terms of such correlations between A-A and pp collisions at LHC energies are discussed in Section V. Conclusions are presented in Section VI.
\section{Collision energy and centrality dependence of $\langle p_{T} \rangle/\sqrt{\langle dN/dy \rangle/S_{\perp}}$; geometrical scaling}
\begin{figure} [htbp]
\includegraphics[width=1.0\linewidth]{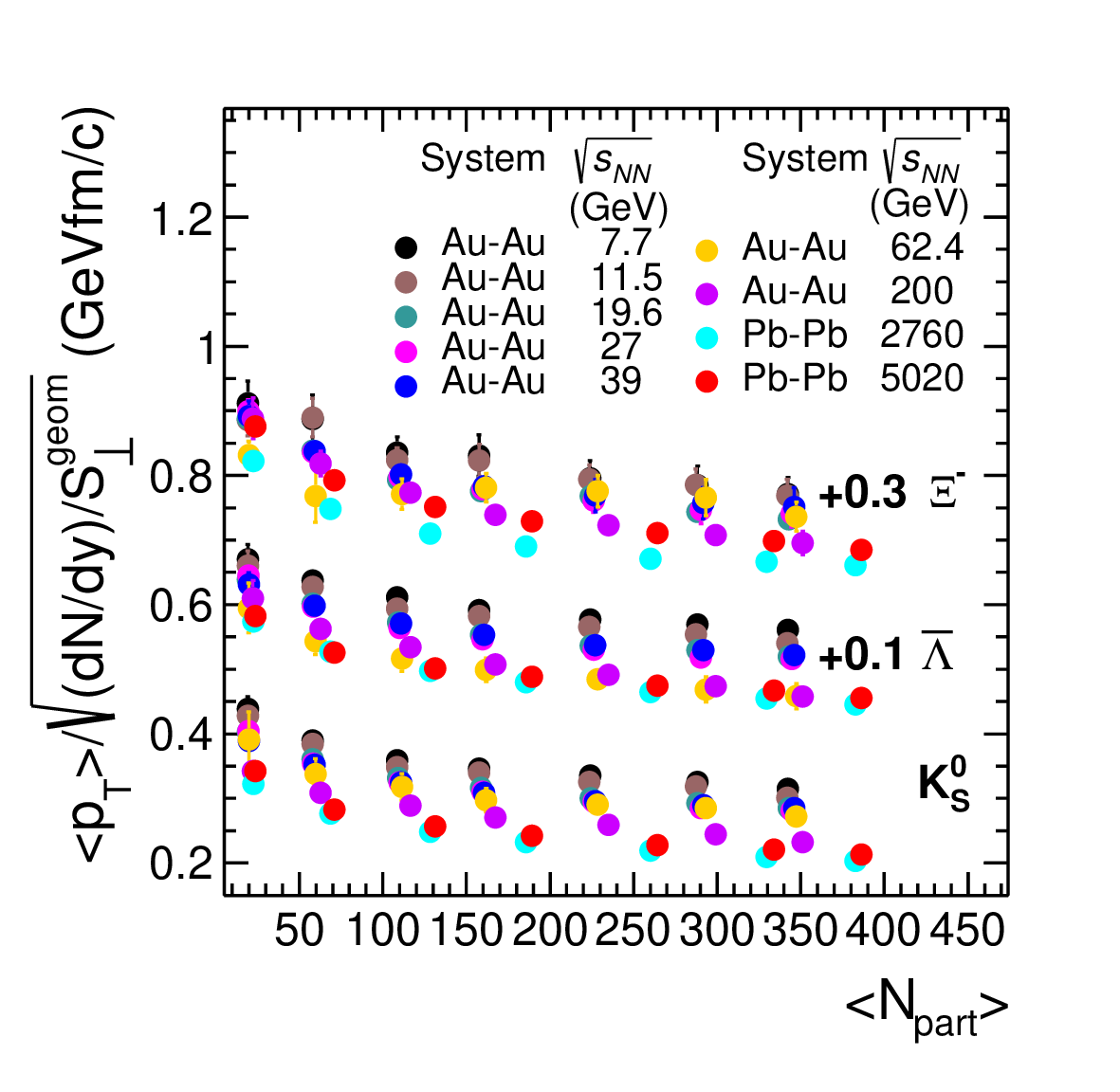}
\caption{$\langle N_{part} \rangle$ dependence of $\langle p_T \rangle/\sqrt{\langle dN/dy \rangle/S_{\perp}^{geom}}$ 
 for ${K^0_S}$, $\bar{\Lambda}$, $\Xi^{-}$ at different collision energies.}
\label{fig-1}
\end{figure}

\begin{figure} [htbp]
\includegraphics[width=1.0\linewidth]{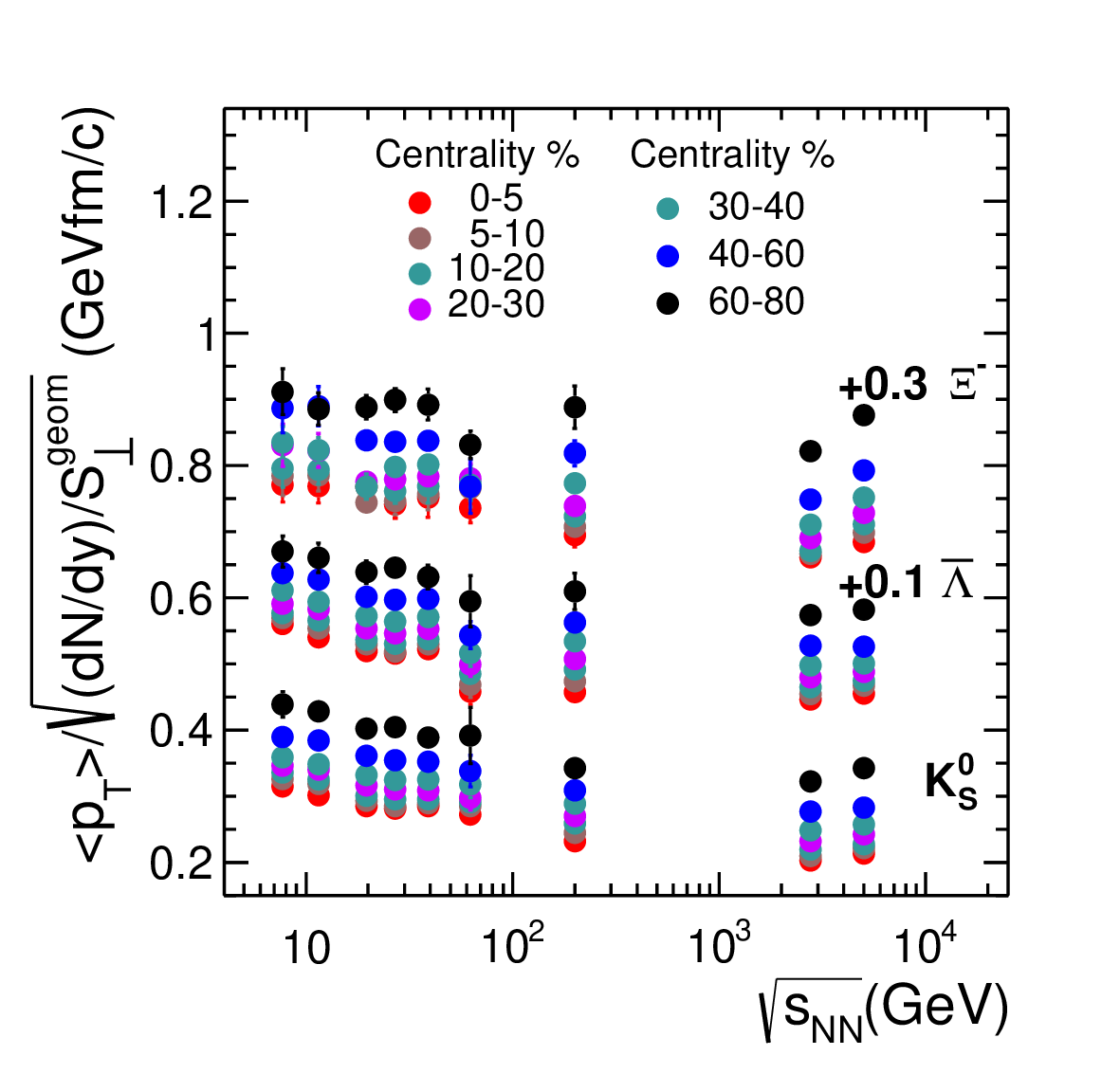}
\caption{Collision energy, $\sqrt{s_{NN}}$, dependence of $\langle p_T \rangle/\sqrt{\langle dN/dy \rangle/S_{\perp}^{geom}}$ for ${K^0_S}$, $\bar{\Lambda}$, $\Xi^{-}$ at different centralities.}
\label{fig-2}
\end{figure}

Previous papers \cite{petro1,petro3} confirmed the model prediction based on local parton-hadron duality (LPHD) \cite{dok1} and dimensionality argument \cite{levin,lappi} concerning the behaviour
 of the ratio between the average transverse momentum
and the square root of the total particle multiplicity per unit 
 rapidity and unit of the colliding nuclei transverse 
overlap area ($\langle p_{T} \rangle/\sqrt{\langle dN/dy \rangle/S_{\perp}}$) as a function of centrality and collision energy for pions, kaons and protons.
 The results of similar studies in the case of strange hadrons ${K^0_S}$, $\bar{\Lambda}$, $\Xi^{-}$ are presented in Figure \ref{fig-1} as a function of
 centrality (number of participating nucleons $\langle N_{part} \rangle$), for different collision energies and in Figure \ref{fig-2} as a function of collision energy, 
 $\sqrt{s_{NN}}$, for different centralities. 
 Similar with the trends shown for pions, kaons and protons \cite{petro1}, a decrease of 
 $\langle p_{T} \rangle/\sqrt{\langle dN/dy \rangle/S_{\perp}}$ with centrality at all collision energies and as a function of collision energy at a given centrality is evidenced. $\langle p_{T} \rangle/\sqrt{\langle dN/dy \rangle/S_{\perp}}$ decreases from $\sqrt{s_{NN}}$ = 7.7 GeV up to 200 GeV, the values corresponding to LHC energies being slightly smaller than the ones corresponding to the top RHIC energy.

  The geometrical transverse overlap area ($S_{\perp}^{geom}$) of the two colliding nuclei for a given
incident energy and centrality was estimated on the basis of a Glauber Monte-Carlo (MC) approach \cite{Glauber55,Franco66,Miller07,Glis} 
as explained in \cite{petro3}  where the calculated values were compiled. These values are used in the present paper.
The total final state particle multiplicity per unit rapidity, $dN/dy$, a measure of the entropy produced in the collision, was estimated according to:
 \begin{widetext}
\begin{equation}\label{eq0}
\frac{dN}{dy}\simeq \frac{3}{2}\frac{dN}{dy}^{(\pi^+ + \pi^-)}+ 
2\frac{dN}{dy}^{(p+\bar{p}, \Xi^- +\bar{\Xi}^+)}  +
\frac{dN}{dy}^{(K^++K^-, \Lambda + \bar{\Lambda}, \Omega^- + \bar{\Omega}^+)} + 2\frac{dN}{dy}^{K^{0}_{S}}
\end{equation}
\end{widetext}
 The experimental data in terms of transverse momentum spectra, yields and average transverse momenta were taken from: \cite{Chatt,Chung} (AGS and SPS), \cite {Adam,STAR4} (Au-Au, Beam Energy Scan (BES)), \cite{STAR1,STAR5} (Au-Au at $\sqrt{s_{NN}}$ = 62.4 GeV), \cite{STAR1,STAR6,STAR7} (Au-Au at $\sqrt{s_{NN}}$ = 200 GeV), \cite{ALICE2,ALICE3,ALICE41,*ALICE42} (Pb-Pb at $\sqrt{s_{NN}}$ = 2.76 TeV) and \cite{ALICE5,ALICE6,ALICE7,ALICE8,balbino} (Pb-Pb at $\sqrt{s_{NN}}$ = 5.02 TeV).
 Eq. \ref{eq0} was applied by taking into consideration all the measured particles and 
anti-particles depending, 
from system to system and case to case, on the most complete published 
experimental information.

 \begin{figure} [htbp]
\includegraphics[width=1.0\linewidth]{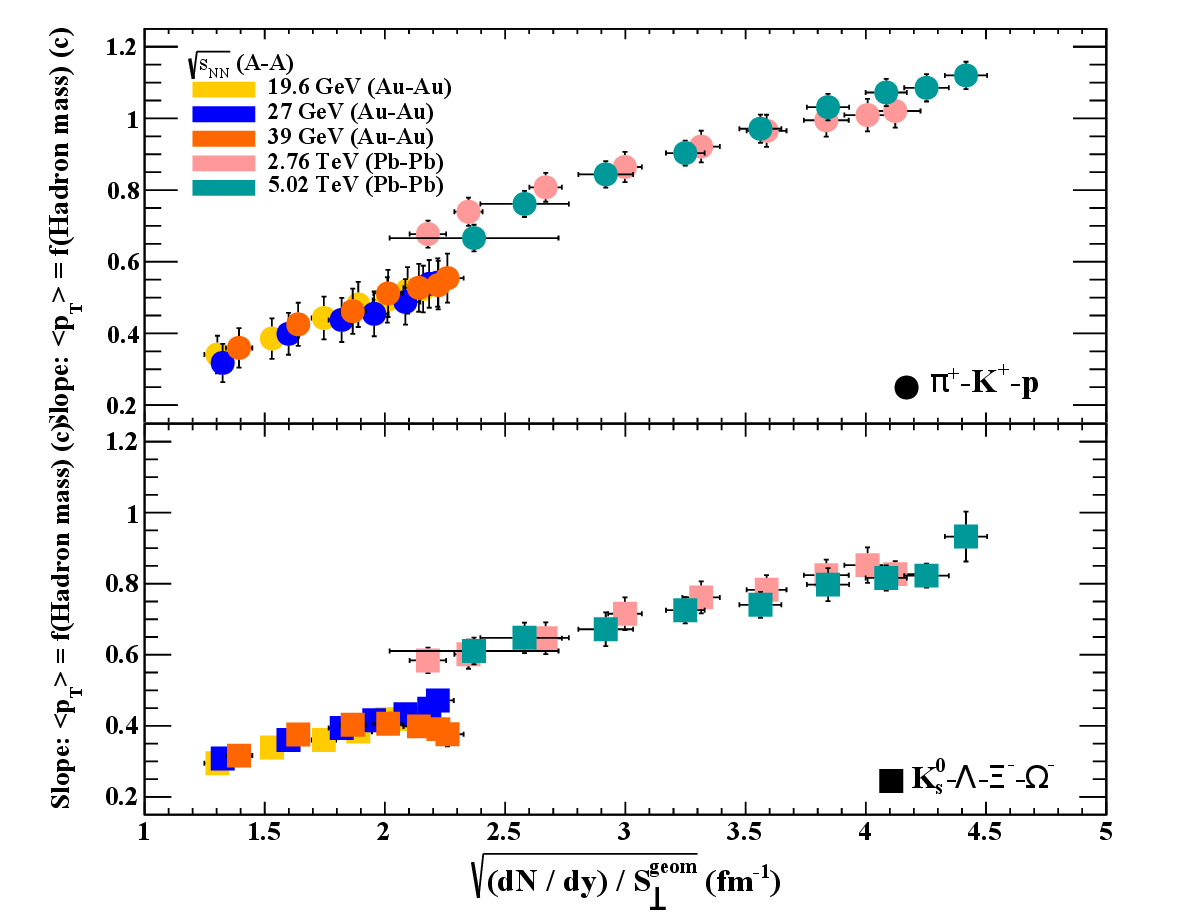}
\caption{$\sqrt{\langle dN/dy \rangle/S_{\perp}}$ dependence of the slope obtained from the linear fit of the 
$\langle p_{T} \rangle$-hadron mass correlation for $\pi^{+}$, $K^{+}$ and p - top and for $K^{0}_{S}$, $\Lambda$, 
$\Xi^{-}$ and $\Omega^{-}$ - bottom.}
\label{fig-3}
\end{figure} 
\begin{figure} [htbp]
\includegraphics[width=1.0\linewidth]{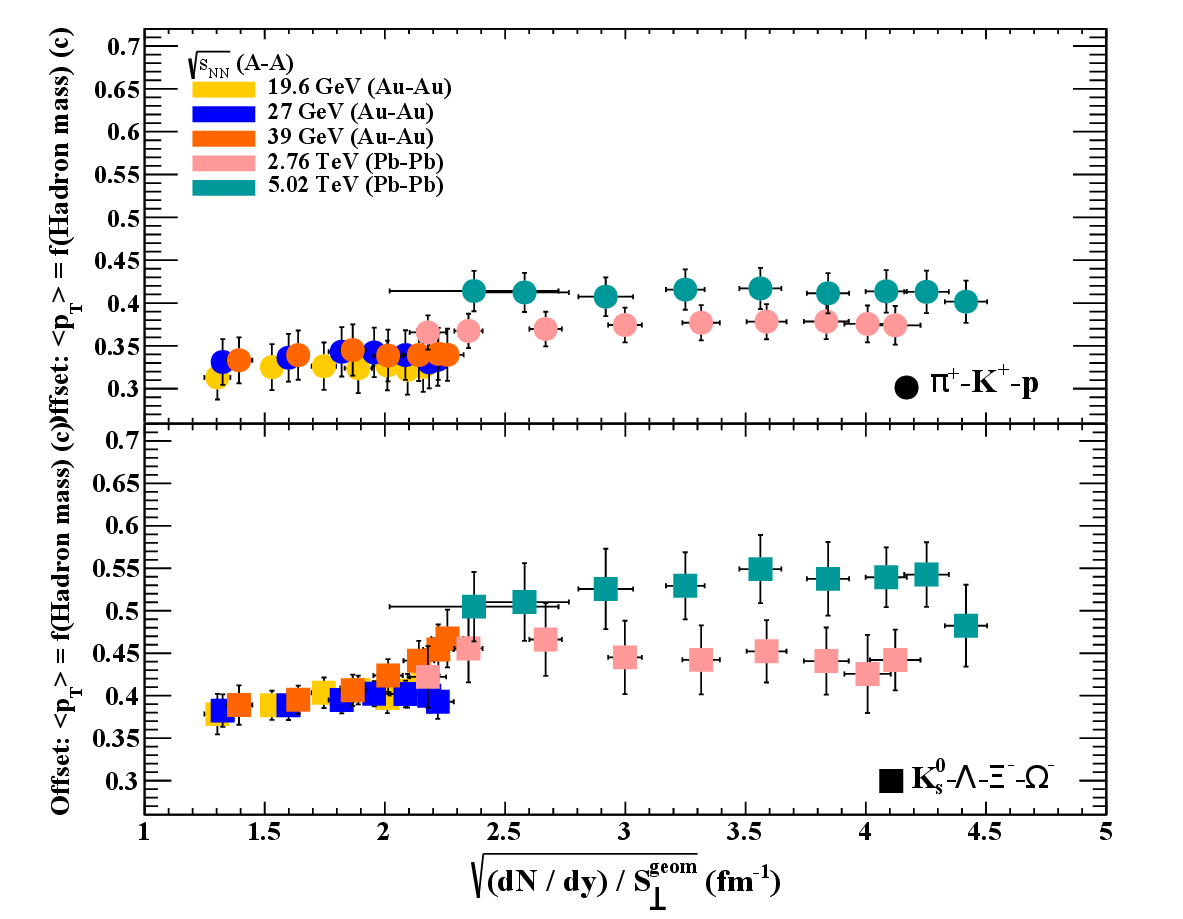}
\caption{$\sqrt{\langle dN/dy \rangle/S_{\perp}}$ dependence 
of the offset obtained from the linear fit of the $\langle p_{T} \rangle$ - hadron mass correlation for $\pi^{+}$, $K^{+}$ and p - top and for 
$K^0_S$, $\Lambda$, $\Xi^{-}$ and $\Omega^{-}$ - bottom.}
\label{fig-4}
\end{figure} 
\begin{figure} [htbp]
\includegraphics[width=1.0\linewidth]{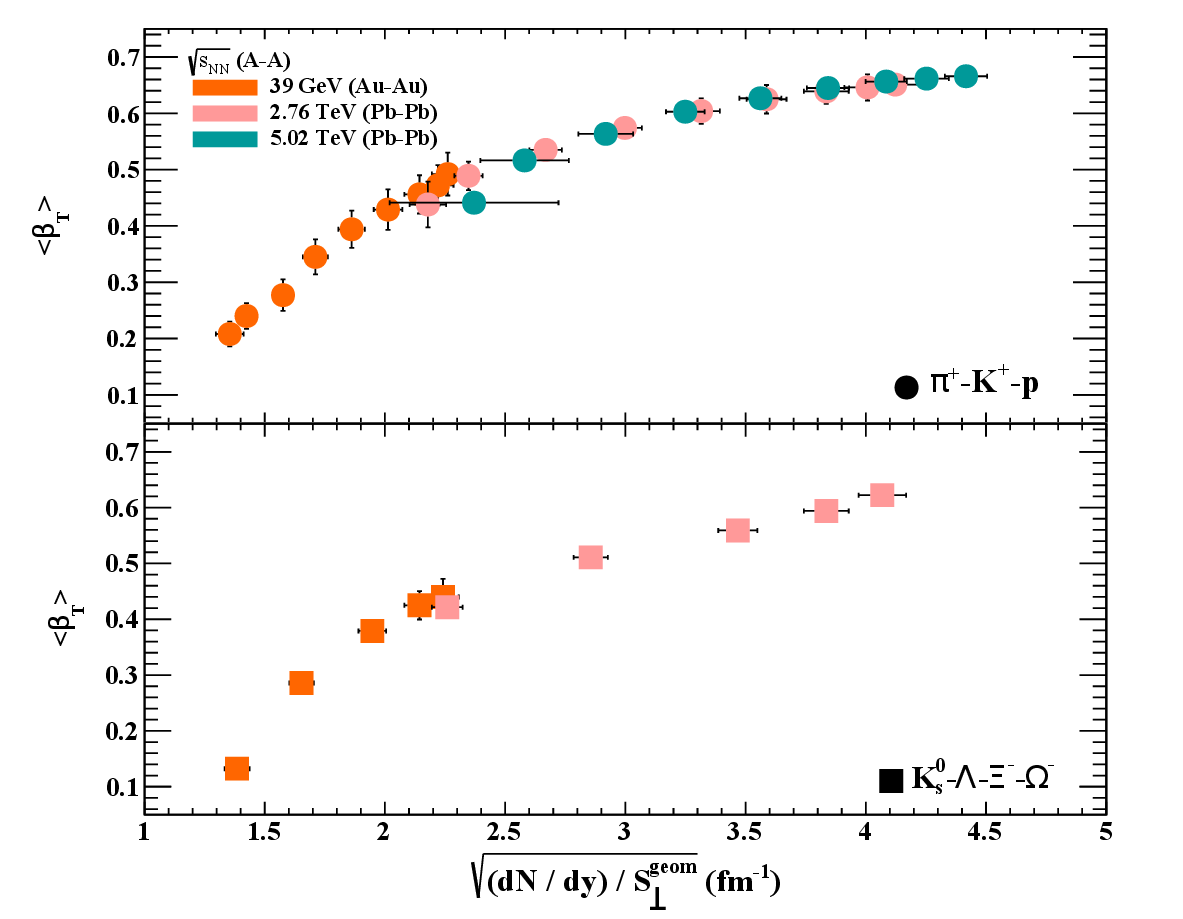}
\caption{$\sqrt{\langle dN/dy \rangle/S_{\perp}}$ dependence of $\langle \beta_{T} \rangle$  obtained from
the BGBW simultaneous fits of $p_{T}$ spectra  for $\pi^{+}$, $K^{+}$ and p - top and for $K^{0}_{S}$, $\Lambda$, $\Xi^{-}$ and 
$\Omega^{-}$ - bottom.} 
\label{fig-5}
\end{figure}  
\begin{figure} [htbp]
\includegraphics[width=1.0\linewidth]{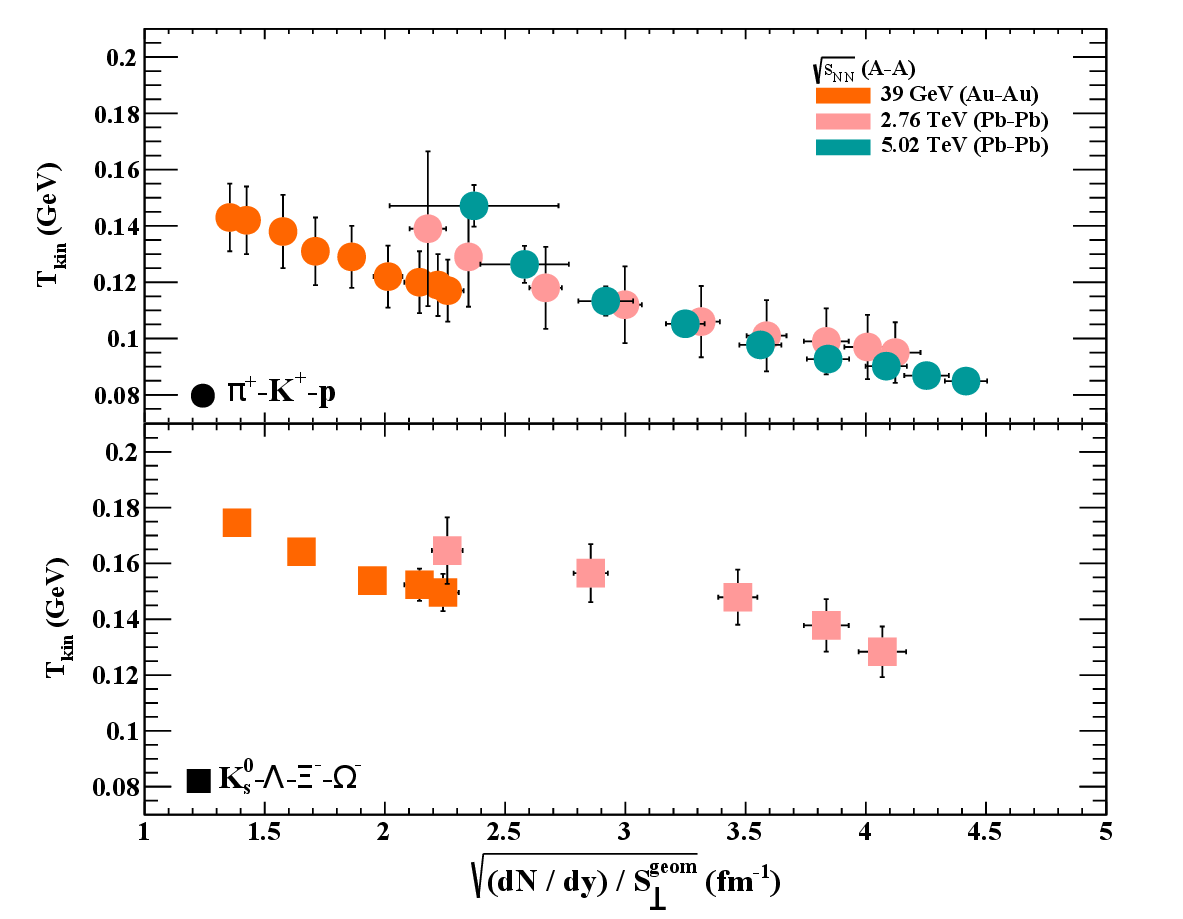}
\caption{$\sqrt{\langle dN/dy \rangle/S_{\perp}}$ dependence of $T_{kin}$  obtained from
the BGBW simultaneous fits of $p_{T}$ spectra for $\pi^{+}$, $K^{+}$ and p - top and for $K^{0}_{S}$, $\Lambda$, $\Xi^{-}$ and 
$\Omega^{-}$ - bottom.}
\label{fig-6}
\end{figure}  
 Results of geometrical scaling, in terms of estimated slopes and offsets  of the linear $\langle p_{T} \rangle$-hadron mass dependence
 and of average transverse flow velocity
 ($\langle \beta_{T} \rangle$) and kinetic freeze-out temperature ($T_{kin}$) parameters from the Boltzmann-Gibbs Blast Wave (BGBW) fits of $p_{T}$ spectra, in relation to the
 geometrical variable $\sqrt{\langle dN/dy \rangle/S_{\perp}}$ and comparison with pp collisions at
 LHC energies for pions, kaons and protons were reported
 in \cite{petro3} and for $K^0_S$, $\Lambda$, $\Xi^{-}$ and $\Omega^{-}$ in \cite{petro4}. The similarities
and differences in terms of these correlations between $\pi^{+}$, $K^{+}$, p and ${K^0_S}$, $\Lambda$, $\Xi^{-}$ and 
$\Omega^{-}$
 can be followed in Figures \ref{fig-3} - \ref{fig-6}.
 Qualitatively, there are similarities between the two categories of hadrons, i.e. a linear dependence on
$\sqrt{\langle dN/dy \rangle/S_{\perp}}$
of the slopes of $\langle p_{T} \rangle$ as a function of 
hadron mass and a constant value of the corresponding offsets, separately for RHIC and LHC energies. 
Quantitatively, there are differences, i.e. the slopes corresponding to
$\pi^{+}$, $K^{+}$ and p are larger than the ones corresponding to $K^0_S, \Lambda, \Xi^{-}$ and $\Omega^{-}$ while the offsets are larger for strange hadrons relative to $\pi^{+}$, $K^{+}$ and p. This is in agreement with the dynamical freeze out scenario, namely, $K^0_S, \Lambda, \Xi^{-}$ and $\Omega^{-}$, suffering negligible interaction with the hadron
environment, conserve the momentum distribution corresponding to their formation time, while  $\pi^{+}$, $K^{+}$ and p could build up extra expansion via elastic interaction with the hadronic medium, resulting in an increase in the slope of  $\langle p_{T} \rangle$ as a function of mass. The offsets, within the error bars, are constant as a function of $\sqrt{\langle dN/dy \rangle/S_{\perp}}$, the values at LHC energies being larger than at RHIC energies.  Within the string percolation scenario, inspired by color glass condensate (CGC) \cite{Bir, Dias} such a difference could be explained by a higher percolation probability at LHC energies relative to the RHIC ones, at LHC energies the gluon density per unit of rapidity and unit overlap area being by a factor 3-4 larger than at the top RHIC energy \cite{Muel}. 
Simultaneous fits of the $p_{T}$ spectra of $\pi^{+}$, $K^{+}$ and p and separately of $K^0_S$, $\Lambda$, $\Xi^{-}$ and $\Omega^{-}$ using the BGBW expression inspired by hydrodynamical models \cite{Schneder} were performed
and the values of the fit parameters namely, the average transverse expansion velocity ($\langle \beta_{T} \rangle$) and 
the kinetic freeze-out temperature ($T_{kin}$), are presented in Figures \ref{fig-5} and \ref{fig-6}, respectively.
$\langle \beta_{T} \rangle$ scales as a 
function of $\sqrt{\langle dN/dy \rangle/S_{\perp}}$, separately for $\pi^{+}$, $K^{+}$ and p and $K^{0}_{S}$, $\Lambda$, $\Xi^{-}$ and 
$\Omega^{-}$ with lower values for strange hadrons at a given $\sqrt{\langle dN/dy \rangle/S_{\perp}}$, 
in agreement with the trends observed in Figure \ref{fig-3}, having the same explanation. The kinetic freeze-out temperature  decreases with increasing $\sqrt{\langle dN/dy \rangle/S_{\perp}}$, the values for $\pi^{+}$, $K^{+}$ and p being lower than the ones corresponding to $K^{0}_{S}$, $\Lambda$, $\Xi^{-}$ and 
$\Omega^{-}$. For both hadron families,  the extracted values at LHC energies are larger than the corresponding
ones at RHIC energies for a given value of $\sqrt{\langle dN/dy \rangle/S_{\perp}}$.
\section{Entropy density dependence of the ratio of strange hadron yields per unit rapidity to the total particle multiplicity per unit rapidity as a function of quark/antiquark composition, collision energy and centrality}  
\begin{figure} [htbp]
\includegraphics[width=1.\linewidth]{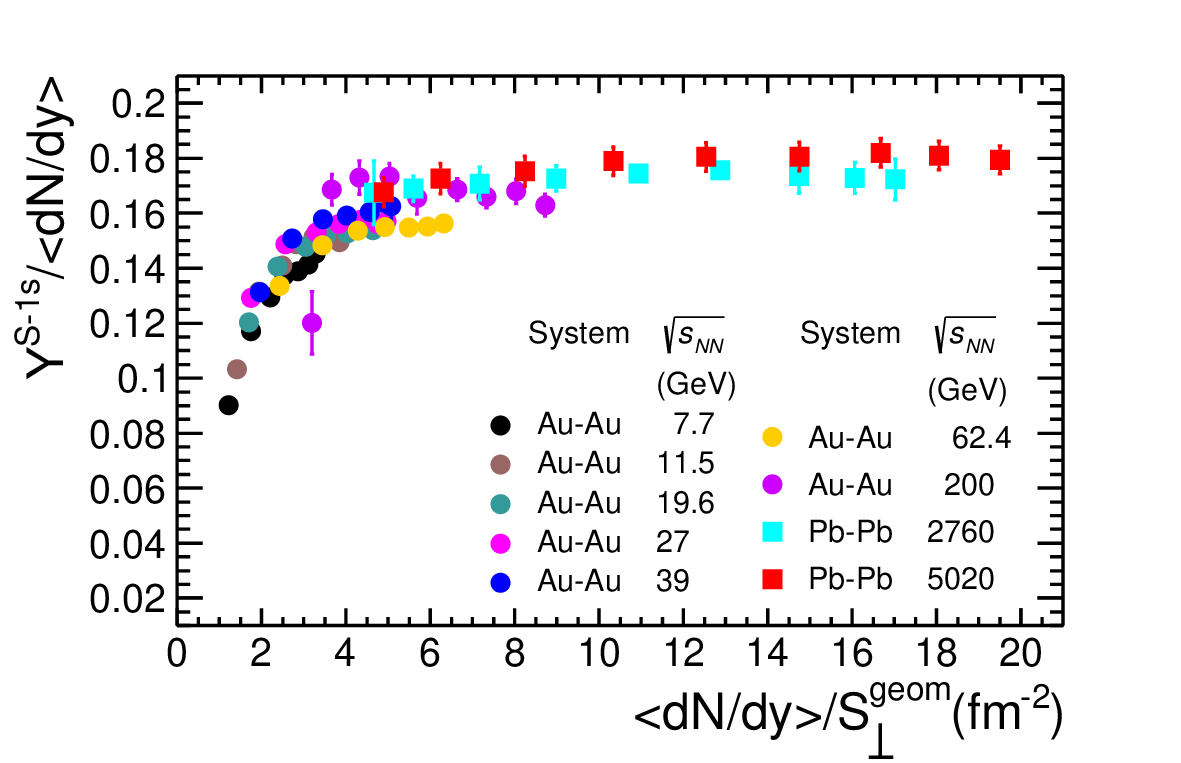}
\caption{The ratio of the single-strange hadron+anti-hadron yield per unit rapidity (Eq. \ref{eq2}) to
the total particle multiplicity per unit rapidity as a function of entropy density from
$\sqrt{s_{NN}}$ = 7.7 GeV up to 5.02 TeV.
}
\label{fig-7}
\end{figure}

\begin{figure} [htbp]
\includegraphics[width=1.\linewidth]{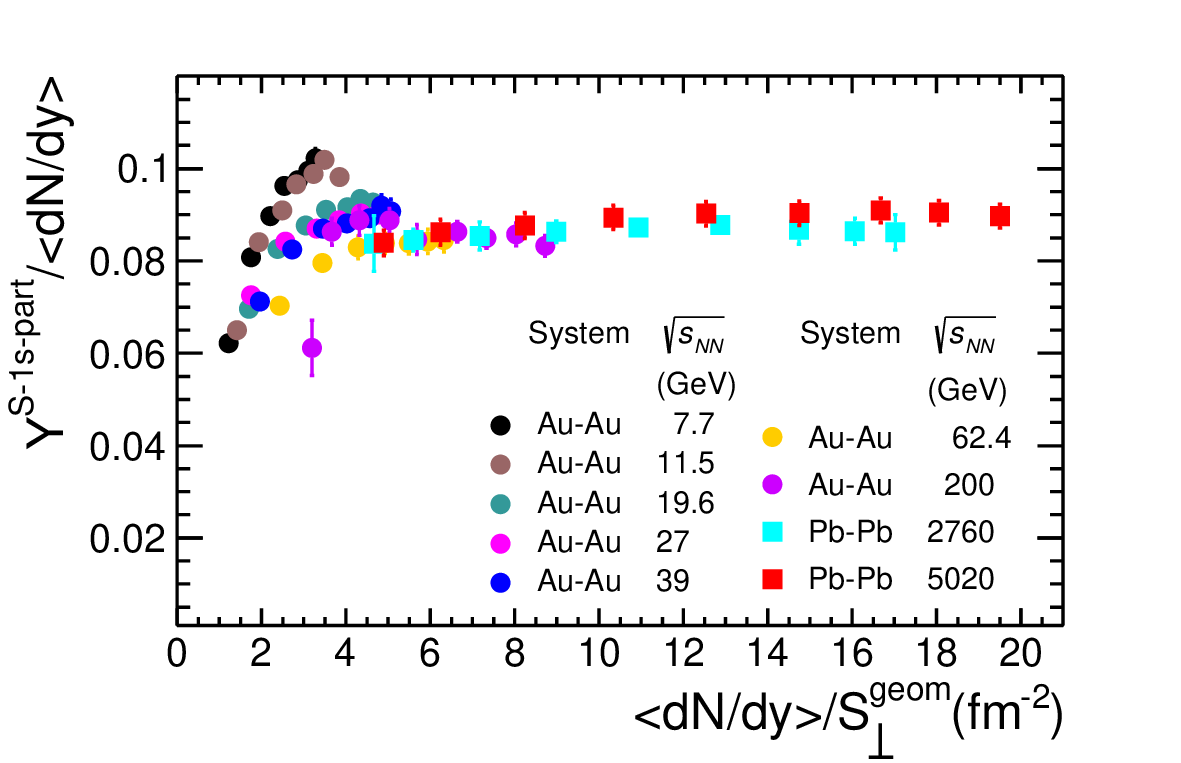}
\caption{The ratio of the single-strange hadron yield per unit rapidity to
the total particle multiplicity per unit rapidity as a function of entropy density from
$\sqrt{s_{NN}}$ = 7.7 GeV up to 5.02 TeV.
}
\label{fig-8}
\end{figure}

\begin{figure} [htbp]
\includegraphics[width=1.\linewidth]{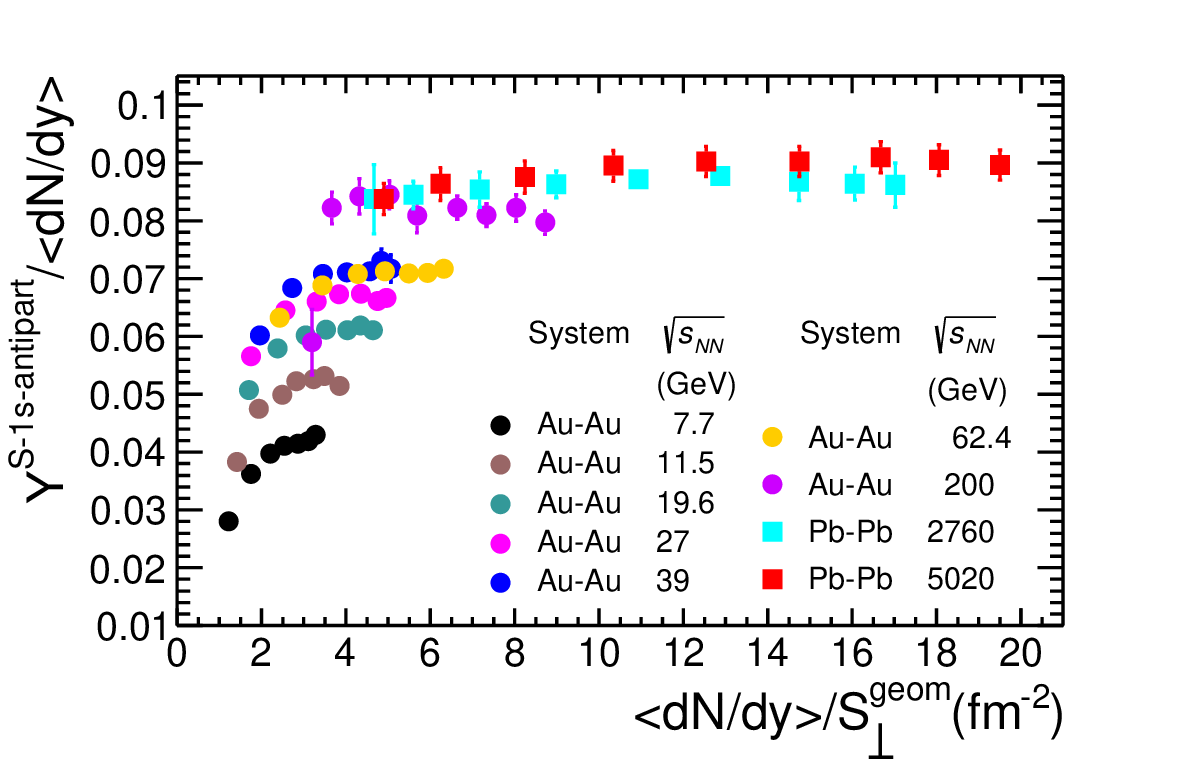}
\caption{The ratio of the single-strange anti-hadron yield per unit rapidity  to
the total particle multiplicity per unit rapidity as a function of entropy density from
$\sqrt{s_{NN}}$ = 7.7 GeV up to 5.02 TeV.
}
\label{fig-9}
\end{figure}

\begin{figure} [htbp]
\includegraphics[width=1.\linewidth]{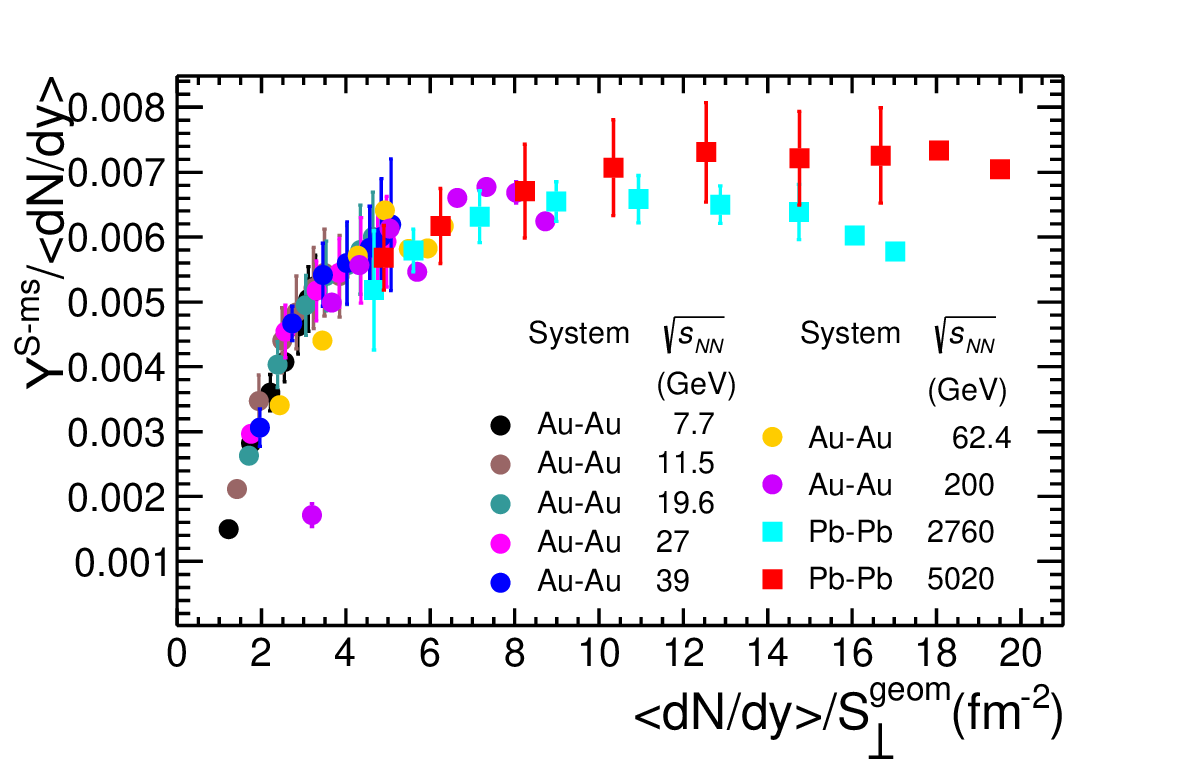}
\caption{The ratio of the multi-strange hadron+anti-hadron yield per unit rapidity (Eq. \ref{eq3}) to
the total particle multiplicity per unit rapidity as a function of entropy density from
$\sqrt{s_{NN}}$ = 7.7 GeV up to 5.02 TeV.
}
\label{fig-10}
\end{figure}

\begin{figure} [htbp]
\includegraphics[width=1.\linewidth]{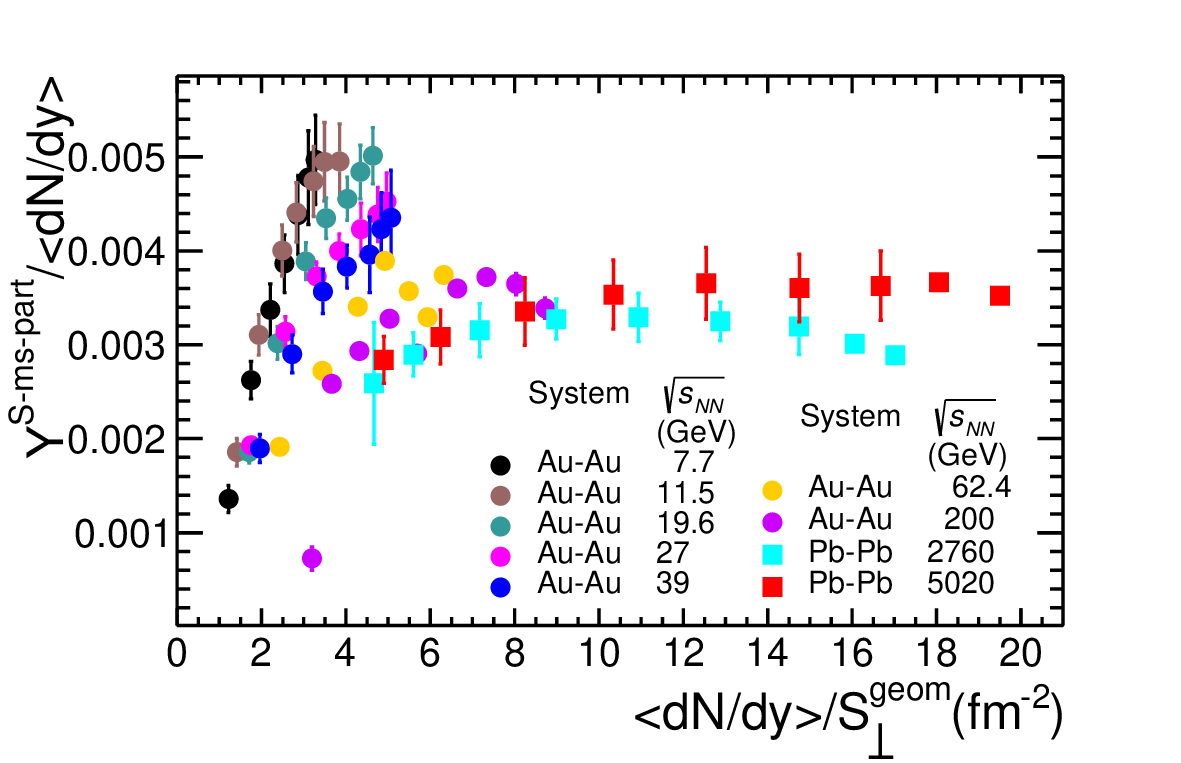}
\caption{The ratio of the multi-strange hadron yield per unit rapidity to
the total particle multiplicity per unit rapidity as a function of entropy density from
$\sqrt{s_{NN}}$ = 7.7 GeV up to 5.02 TeV.
}
\label{fig-11}
\end{figure}

\begin{figure} [htbp]
\includegraphics[width=1.\linewidth]{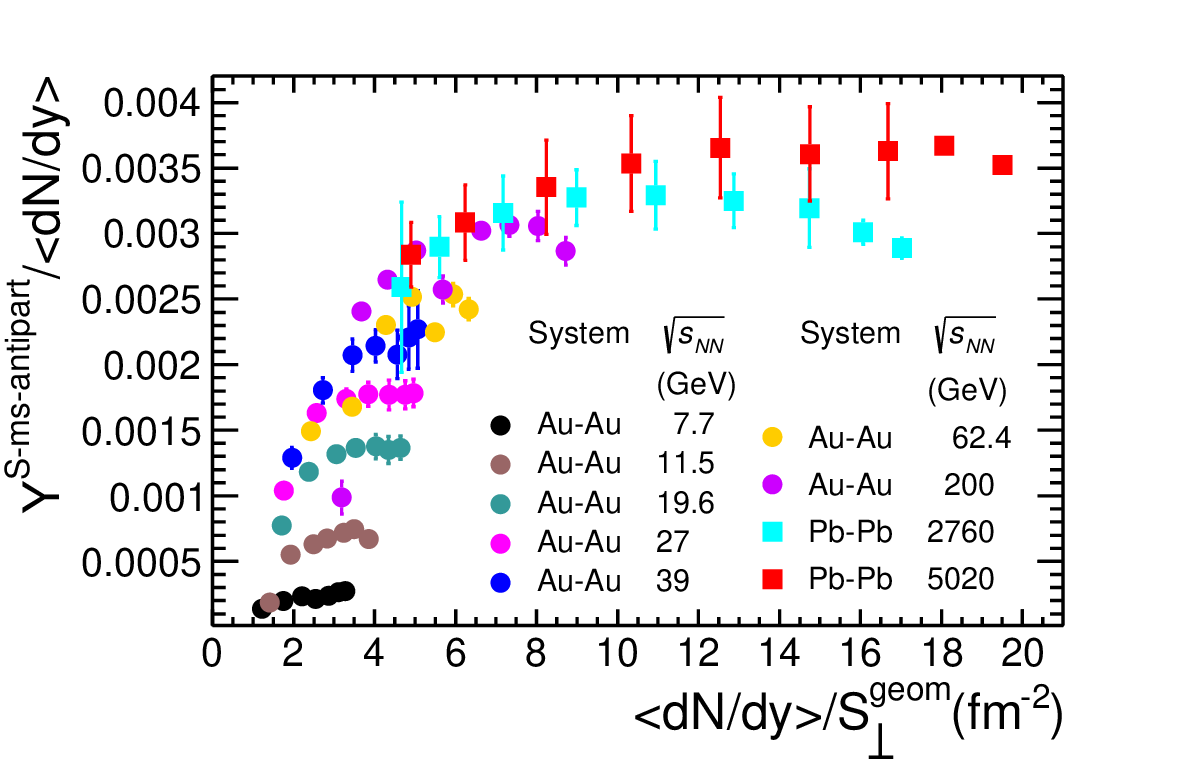}
\caption{The ratio of the multi-strange anti-hadron yield per unit rapidity  to
the total particle multiplicity per unit rapidity as a function of entropy density from
$\sqrt{s_{NN}}$ = 7.7 GeV up to 5.02 TeV.
}
\label{fig-12}
\end{figure}

Since the entropy and strangeness are conserved during the confinement process, the ratio between the
strange hadron yield and entropy has been proposed as diagnostic tool for deconfinement \cite{rafelski3,rafelski2}. The entropy density
per unit transverse overlap area, assuming an isentropic expansion, is proportional to the total particle
multiplicity per unit rapidity and unit transverse overlap area. In order to compare different collision systems,
Au-Au at RHIC energies, Pb-Pb and pp at LHC energies, based on the present experimental information,
the dependence of the ratio between the strange hadron yield  per
unit rapidity and the total particle multiplicity per unit rapidity on the entropy density ($\langle dN/dy \rangle/S_{\perp}$) was scrutinized.   
Such
correlations are investigated separately for single- and
multi-strange hadrons. 
The  total single- ($Y^{S-1s}$) and multi- ($Y^{S-ms}$) strange hadron yields per unit rapidity, are estimated from the existing experimental data with:  
\begin{widetext}
\begin{equation}\label{eq2}
Y^{S-1s}=\frac{dN}{dy}^{S-1s}=\frac{dN}{dy}^{(K^++K^-)}+2\frac{dN}{dy}^{K^{0}_{S}}+\frac{dN}{dy}^{(\Lambda + \bar{\Lambda})}
\end{equation}
\begin{equation}\label{eq3}
Y^{S-ms}=\frac{dN}{dy}^{S-ms}=2\frac{dN}{dy}^{(\Xi^- +\bar{\Xi}^+)}  +
\frac{dN}{dy}^{(\Omega^- + \bar{\Omega}^+)} 
\end{equation}
\end{widetext}
By examining Figure \ref{fig-7}, where the ratio of the total single-strange hadron yield per unit rapidity (Eq. \ref{eq2}) to the total particle multiplicity per unit rapidity as a function of entropy density is presented,  one can conclude that below $\sqrt{s_{NN}}$ = 62.4 GeV a scaling is evidenced, 
 $Y^{S-1s}/\langle dN/dy \rangle$
increasing with the entropy density. 
A similar trend is observed at $\sqrt{s_{NN}}$ = 62.4 GeV
with a systematic lower value of $Y^{S-1s}/\langle dN/dy \rangle$
for a given entropy density. Within the error bars, the results at  $\sqrt{s_{NN}}$ = 200 GeV
seem to be in line with the LHC values, $Y^{S-1s}/\langle dN/dy \rangle$ 
being almost independent on the entropy density. 
Early theoretical studies of strangeness in relativistic
heavy ion collisions \cite{koch1,koch2} have concluded that especially strange anti-hadrons are sensitive probes for deconfinement. Therefore,
the same type of representations, separately for single-strange hadrons and anti-hadrons  are presented in Figures \ref{fig-8} and \ref{fig-9}, respectively. Figure \ref{fig-8}  clearly shows that
$Y^{S-1s-part}/\langle dN/dy \rangle$
for single-strange hadrons increases with entropy
density for a given collision energy and decreases with the
collision energy. This trend is a consequence of in-medium effects in the presence of baryonic matter in
the mid-rapidity region which decreases with the collision energy as the net baryonic matter at mid-rapidity approaches a negligible value at the top RHIC energy, the collision becoming almost transparent \cite{Adam}. 
The trend for the single-strange anti-hadrons presented in Figure \ref{fig-9}  is completely
different, i.e. $Y^{S-1s-antipart}/\langle dN/dy \rangle$ increases with the collision energy up to $\sqrt{s_{NN}}$ = 200 GeV and
for a given collision energy increases with entropy density bending towards larger entropy density. The collision energy and entropy density dependence of single-strange anti-hadrons is indicative of their production once the creation of pairs of quarks and antiquarks becomes possible. The tendency towards a levelling off at large entropy density is the consequence of the surrounding hadronic matter which increases towards central collisions or larger $\langle dN/dy \rangle/S_{\perp}$. At LHC energies where the  mechanism of creation of pairs of quarks and antiquarks becomes dominant and most of the fireball reaches a deconfined phase, there is no difference between the behaviour of single-strange hadrons and single-strange anti-hadrons and the dependence of  $Y^{S}/\langle dN/dy \rangle$ on the entropy density becomes negligible.
 The results from 
$\sqrt{s_{NN}}$ = 200 GeV are in-line with the LHC ones at the same entropy density and no dependence on the entropy density can be observed within the experimental errors. The $Y^{S}/\langle dN/dy \rangle$ values corresponding to 
the Pb-Pb collision at $\sqrt{s_{NN}}$ = 5.02 TeV are systematically a bit larger than the ones corresponding to 
the Pb-Pb collision at $\sqrt{s_{NN}}$ = 2.76 TeV. 
 The multi-strange hadron yields are more sensitive to the
production mechanism originating in the fireball, therefore are expected to be more sensitive to the fireball composition. 
Similar to the previous cases, it is worth to look to  $Y^{S-ms}/\langle dN/dy \rangle$ - 
$\langle dN/dy \rangle/S_{\perp}$
 for combined multi-strange hadrons and their anti-partners (Eq. \ref{eq3}). The result is presented in Figure \ref{fig-10}. 
 A rather good scaling can be observed with a steep linear
rise up to $\langle dN/dy \rangle/S_{\perp}$ $\approx$ 3 $fm^{-2}$. At the top RHIC energies,
a transition to the regions corresponding to
LHC energies is observed. 
However, this result is nothing else than a combination of two very different dependences of 
$Y^{S}/\langle dN/dy \rangle$ as a function of
$\langle dN/dy \rangle/S_{\perp}$ for multi-strange hadrons and multi-strange anti-hadrons as can be followed in Figures \ref{fig-11} and \ref{fig-12}. 
For multi-strange hadrons, the considerations made relative to the trends evidenced in Figure \ref{fig-8} are more evident.
The $Y^{S-ms-part}/\langle dN/dy \rangle$ for multi-strange hadrons has a steep
linear increase with the entropy density, the slope decreasing going from 
$\sqrt{s_{NN}}$ = 7.7 GeV up to 39 GeV, at 200 GeV converging towards the values corresponding to
$\sqrt{s_{NN}}$ = 5.02 TeV. At the LHC energy of $\sqrt{s_{NN}}$ = 5.02 TeV, a very small increase of 
$Y^{S-ms-part}/\langle dN/dy \rangle$ as a function of entropy density is observed.
The entropy density dependence of $Y^{S-ms-antipart}/\langle dN/dy \rangle$ for the
multi-strange anti-hadrons presented in Figure \ref{fig-12} is
completely different. $Y^{S-ms-antipart}/\langle dN/dy \rangle$ increases with collision
energy and for a given collision energy increases with the entropy density with a tendency towards a plateau at larger entropy density/larger centrality. 
These results show that an integral representation as those presented in Figures \ref{fig-7} and \ref{fig-10}, in which strange hadrons
and anti-hadrons are summed together, hides all the trends revealed in separate representations, Figures \ref{fig-8}, \ref{fig-9}, \ref{fig-11} and \ref{fig-12}, and
makes it difficult to take it as testing ground for theoretical models. Moreover, because of the much larger yield of single-strange hadrons relative to the multi-strange hadrons one, the behaviour of the established correlations  in the second case is obscured in an integral representation for the total strange hadron yield.
The trends evidenced in Figures \ref{fig-9} and  \ref{fig-12} are very similar with those observed in 
the $\langle dE_{T} /dy \rangle/\langle dN/dy \rangle$ - $\langle dN/dy \rangle/S_{\perp}$ correlation for different collision energies
and different centralities \cite{petro1}. This shows that there is a strong correlation between the strange anti-hadron production and energy density. A similar conclusion was reached by studying the degree of strangeness suppression in hadronic and nuclear collisions \cite{castorina}.
\section{Various aspects related to the $Y^{S}/\langle dN/dy \rangle$ - $\langle dN/dy \rangle/S_{\perp}$ correlation for different quark/antiquark compositions at different centralities}
\begin{figure} [htbp]
\includegraphics[width=0.95\linewidth]{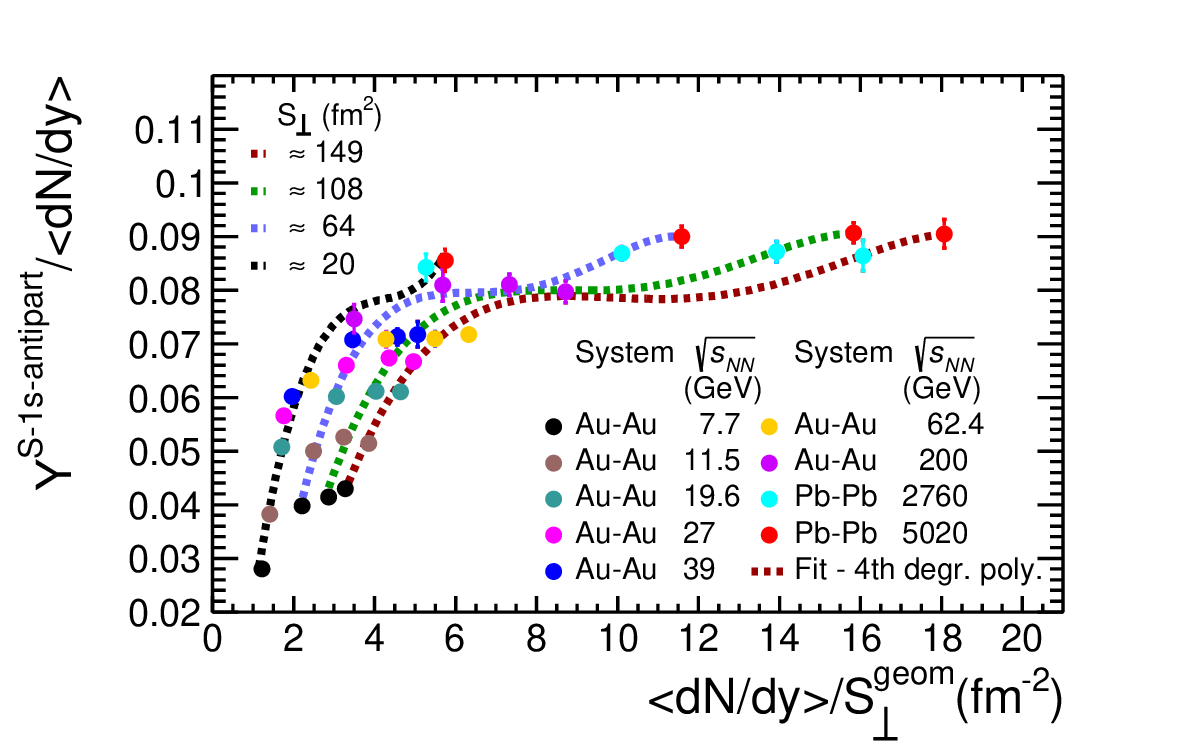}
\caption{The ratio of the single-strange anti-hadron yield per unit rapidity  to
the total particle multiplicity per unit rapidity as a function of entropy density 
for four values of the transverse overlap area. The dashed lines are the result of a fourth degree polynomial fit
used to guide the eye.
}
\label{fig-13}
\end{figure}

\begin{figure} [htbp]
\includegraphics[width=1.\linewidth]{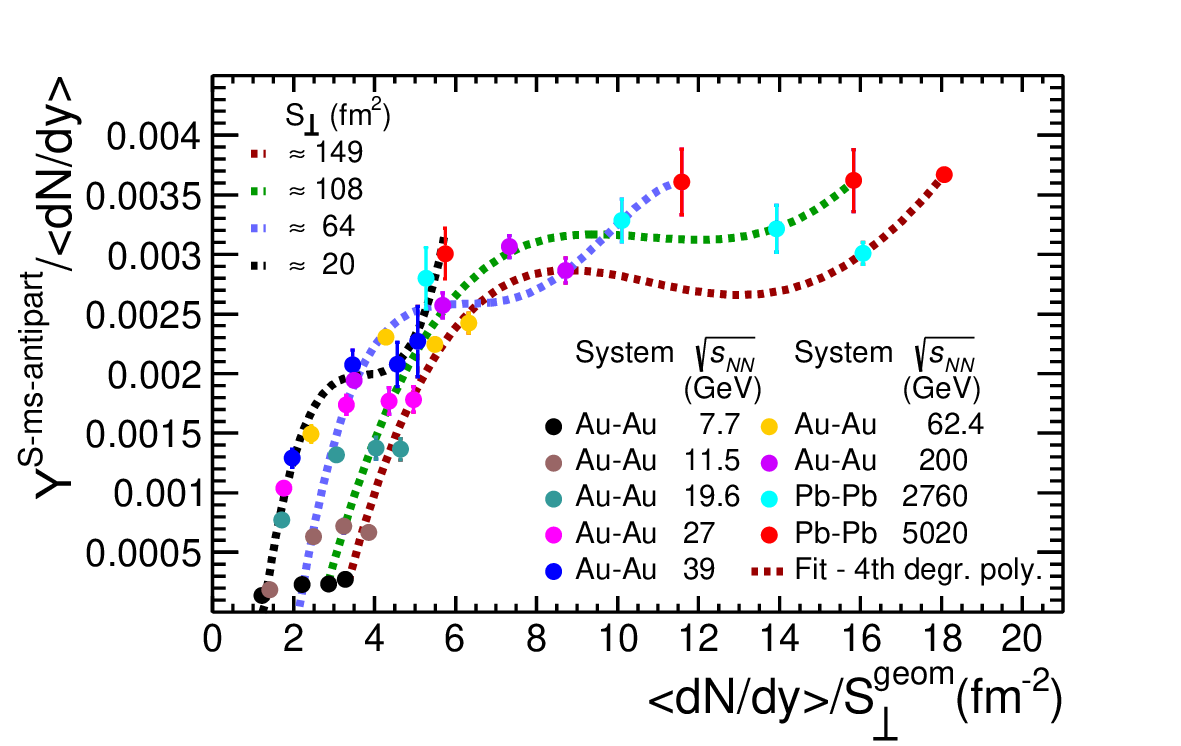}
\caption{The ratio of the multi-strange anti-hadron yield per unit rapidity  to
the total particle multiplicity per unit rapidity as a function of entropy density 
for four values of the transverse overlap area. The dashed lines are the result of a fourth degree polynomial fit
used to guide the eye.
}
\label{fig-14}
\end{figure}

Four different values of the transverse overlap area, namely,  $S_{\perp}$ = 149, 108, 64, 20 $fm^{2}$ have been chosen to inspect the fireball size dependence of the strangeness/entropy-entropy density correlation. The corresponding values of the ratio of the strange anti-hadron yields per unit rapidity  to the total particle multiplicity per unit rapidity were obtained with the interpolation of the 
experimental data measured at different centralities.  The
results are presented in Figures \ref{fig-13} and \ref{fig-14} for single- and multi- strange anti-hadrons, respectively.
A clear dependence on the size of the transverse overlap area is observed. The rise
in $Y^{S-1s-antipart}/\langle dN/dy \rangle$ and $Y^{S-ms-antipart}/\langle dN/dy \rangle$ at RHIC energies becomes steeper and
the range in $\langle dN/dy \rangle/S_{\perp}$, corresponding to a trend close to a plateau, decreases from central to peripheral collisions. The trends resemble those observed in Fig. 5 from \cite{petro1} for the $\langle dE_{T} /dy \rangle/\langle dN/dy \rangle$ - $\langle dN/dy \rangle/S_{\perp}$ correlation. 
The dashed lines represent the result of the fit  of the corresponding
points with a fourth degree polynomial.
At each inflection point of the fit lines,
the corresponding values of the ratio of the yield per unit rapidity to the total particle multiplicity per unit rapidity were estimated and represented in Figures \ref{fig-15} and \ref{fig-16} for single- and multi-strange anti-hadrons, respectively. 
While these values are constant in Figure {\ref{fig-15}, in Figure \ref{fig-16}, which looks like Figure 6 in \cite{petro1}, a slight dependence on the overlap area is observed.
\begin{figure} [htbp]
\includegraphics[width=1.\linewidth]{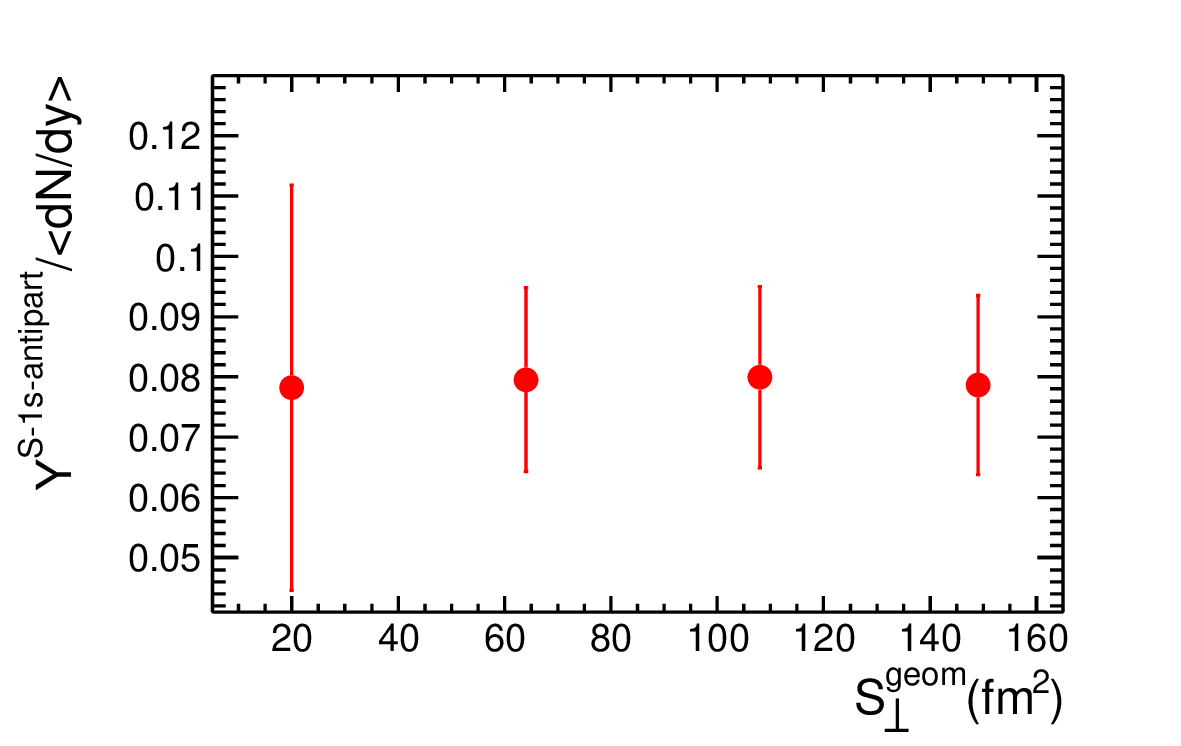}
\caption{$Y^{S-1s-antipart}/\langle dN/dy \rangle$ calculated at the inflection point in
the $Y^{S-1s-antipart}/\langle dN/dy \rangle$ - $\langle dN/dy \rangle/S_{\perp}$ correlation at different collision energies for four values of
$S^{geom}_{\perp}$.
}
\label{fig-15}
\end{figure}  
\begin{figure} [htbp]
\includegraphics[width=1.\linewidth]{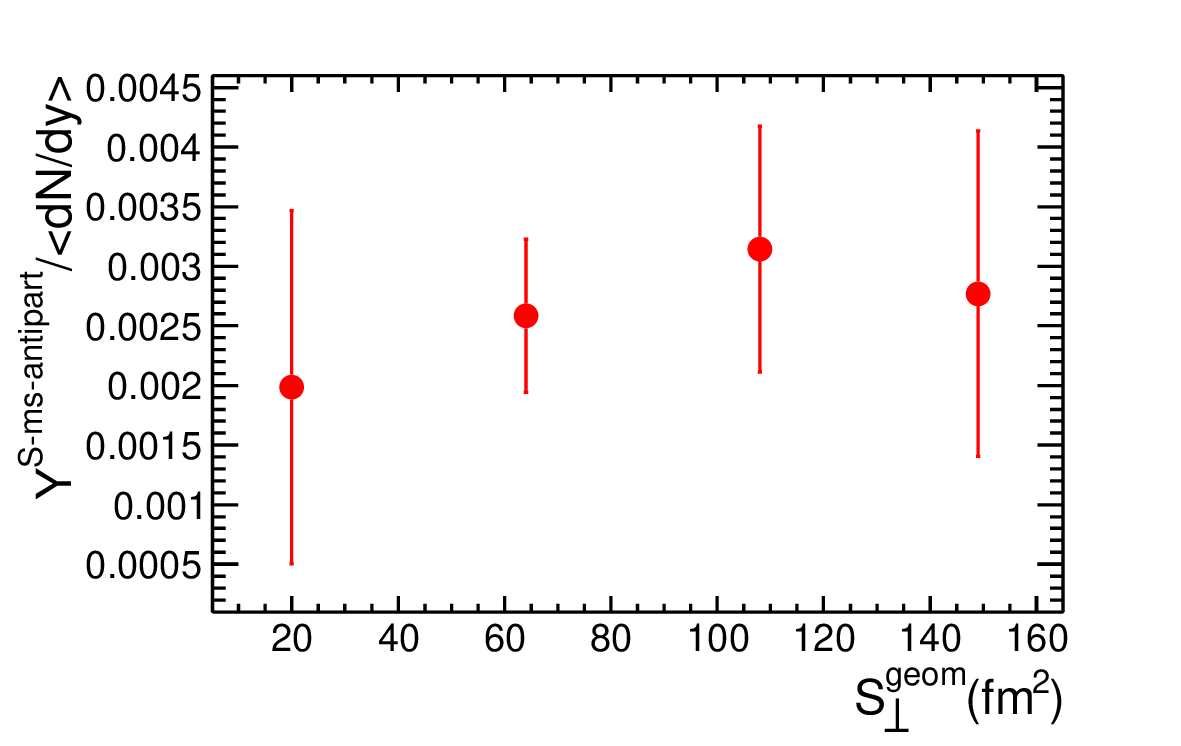}
\caption{$Y^{S-ms-antipart}/\langle dN/dy \rangle$ calculated at the inflection point in
the $Y^{S-ms-antipart}/\langle dN/dy \rangle$ - $\langle dN/dy \rangle/S_{\perp}$ correlation at different collision energies for four values of
$S^{geom}_{\perp}$.
}
\label{fig-16}
\end{figure}  

These considerations and the following ones  can
be better understood based on the collision energy dependence of
$\langle dN/dy \rangle/S_{\perp}$ for different centralities, presented in Figure \ref{fig-17}. 
\begin{figure} [htbp]
\includegraphics[width=1.\linewidth]{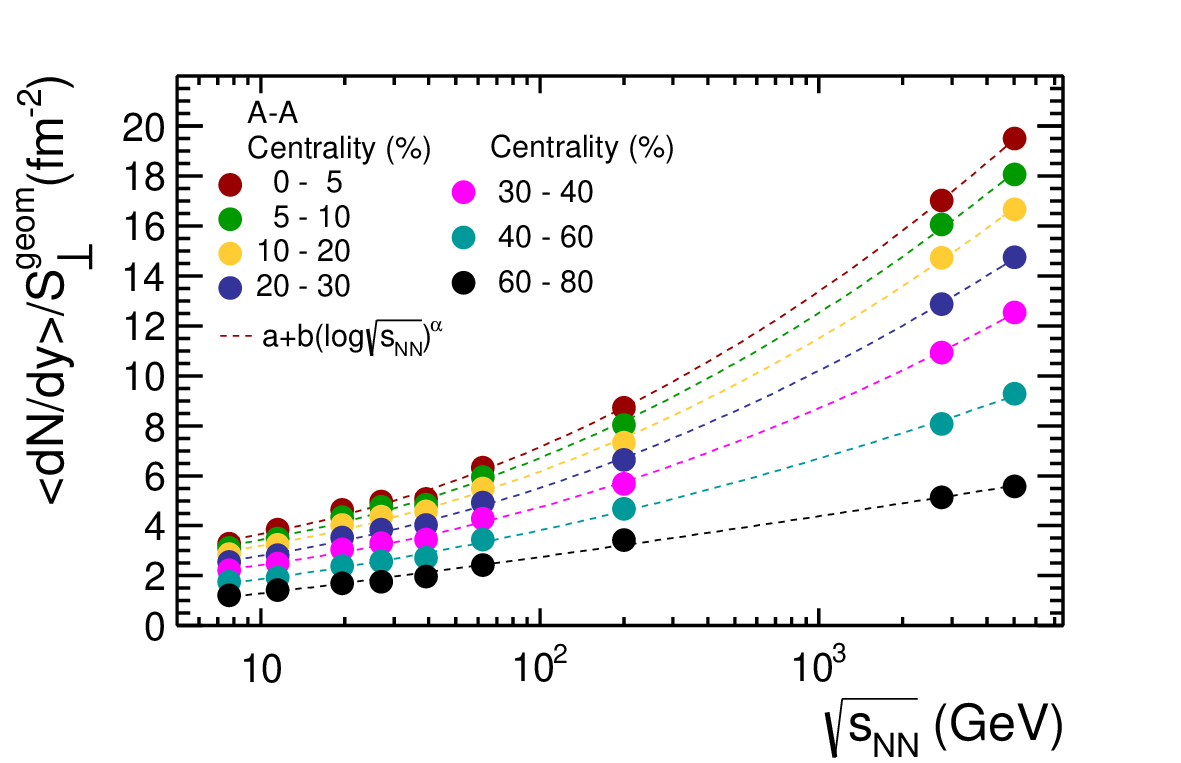}
\caption{Collision energy dependence of $\langle dN/dy \rangle/S_{\perp}$ for different centralities.
}
\label{fig-17}
\end{figure}  
The range in $\langle dN/dy \rangle/S_{\perp}$ as a function of collision energy is increasing from peripheral to central collisions.
The dashed lines are the result of the fit with the formula $\langle dN/dy \rangle/S_{\perp}=a+b(\log{\sqrt{s_{NN}}})^{\alpha}$ and the fit coefficients corresponding to the different centralities are listed in Table \ref{tab1}.
\begin{table*}
\caption{\label{tab1}Coefficients of the fit with the formula from Figure \ref{fig-17}}
%\caption{\label{tab1}}
\begin{ruledtabular}
\begin{tabular}{ c c c c}
 Centrality (\%) & a & b & $\alpha$\\ 
  0 - 5 & 2.6399  $\pm$ 0.2284   & 1.0239 $\pm$ 0.1365 &   1.0700  $\pm$   0.0475 \\	
  5 - 10& 2.4497   $\pm$   0.2338 & 0.9829   $\pm$   0.1413 & 1.0590   $\pm$   0.0511\\  
 10 - 20& 2.3246   $\pm$   0.1970 & 0.8712   $\pm$   0.1176 & 1.0718   $\pm$   0.0481 \\  
 20 - 30& 1.9471   $\pm$   0.1528& 0.8518   $\pm$   0.0944 &  1.0341   $\pm$   0.0392 \\  
 30 - 40& 1.7053   $\pm$   0.1438 & 0.7293   $\pm$   0.0893 & 1.0293   $\pm$   0.0433 \\  
 40 - 60& 1.0456   $\pm$   0.1861 & 0.8217   $\pm$   0.1314 & 0.8767   $\pm$   0.0544\\  
 60 - 80& 0.1873   $\pm$   0.3645 & 1.1015   $\pm$   0.3083 &   0.6086   $\pm$   0.0844
\end{tabular}
\end{ruledtabular}
\end{table*}

In Figures \ref{fig-18} and \ref{fig-19} are represented the ratios of single- and multi-strange hadron (full symbols) and single- and multi-strange anti-hadron (open symbols), respectively, yields per unit rapidity to the total particle multiplicity per unit rapidity as a  function of entropy density, for four values of centrality. The ratio for single- and multi-strange hadrons  relative to anti-hadrons, at lower entropy density, decreases from central to peripheral collisions, becoming similar towards the plateau region. This shows that decreasing the number of participants the single collisions become the main source of strange hadron production.
\begin{figure} [htbp]
\includegraphics[width=1.\linewidth]{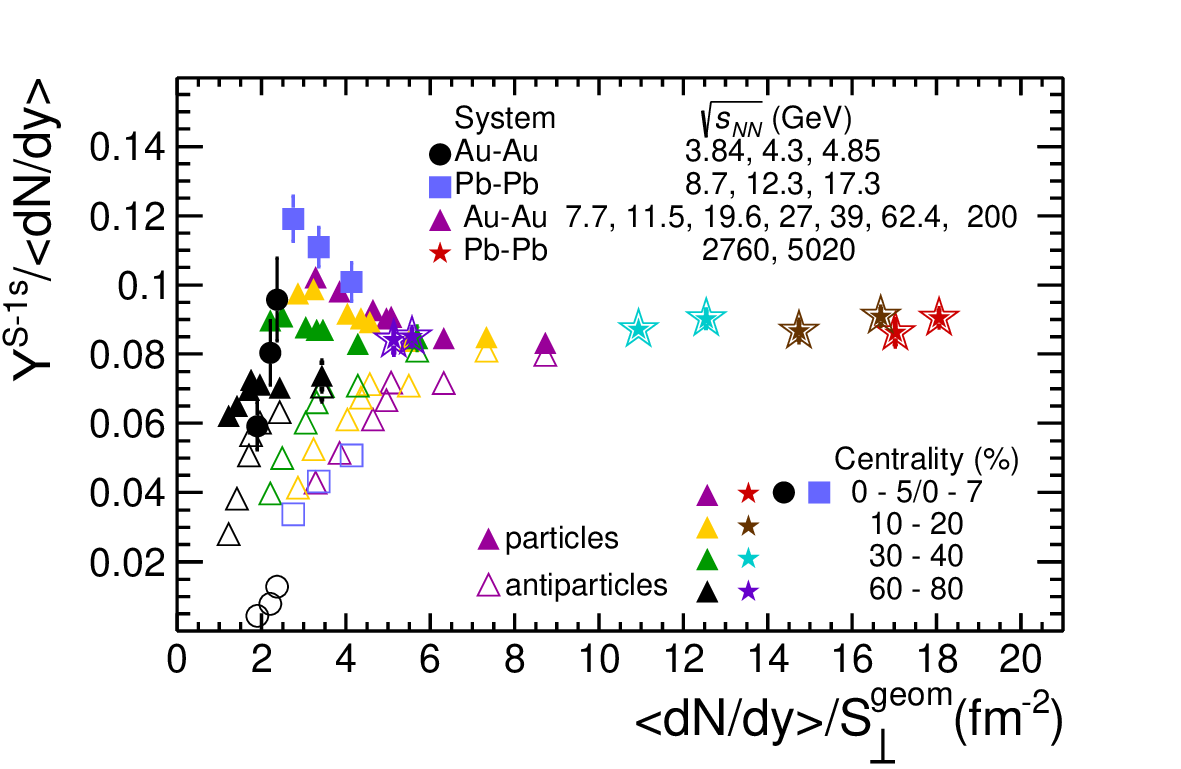}
\caption{The ratio of single-strange hadron (full symbols)
and single-strange anti-hadron (open symbols) yields per unit rapidity to the total particle multiplicity per unit rapidity as a function of entropy density
for four values of centrality.
}
\label{fig-18}
\end{figure}
\begin{figure} [htbp]
\includegraphics[width=1.\linewidth]{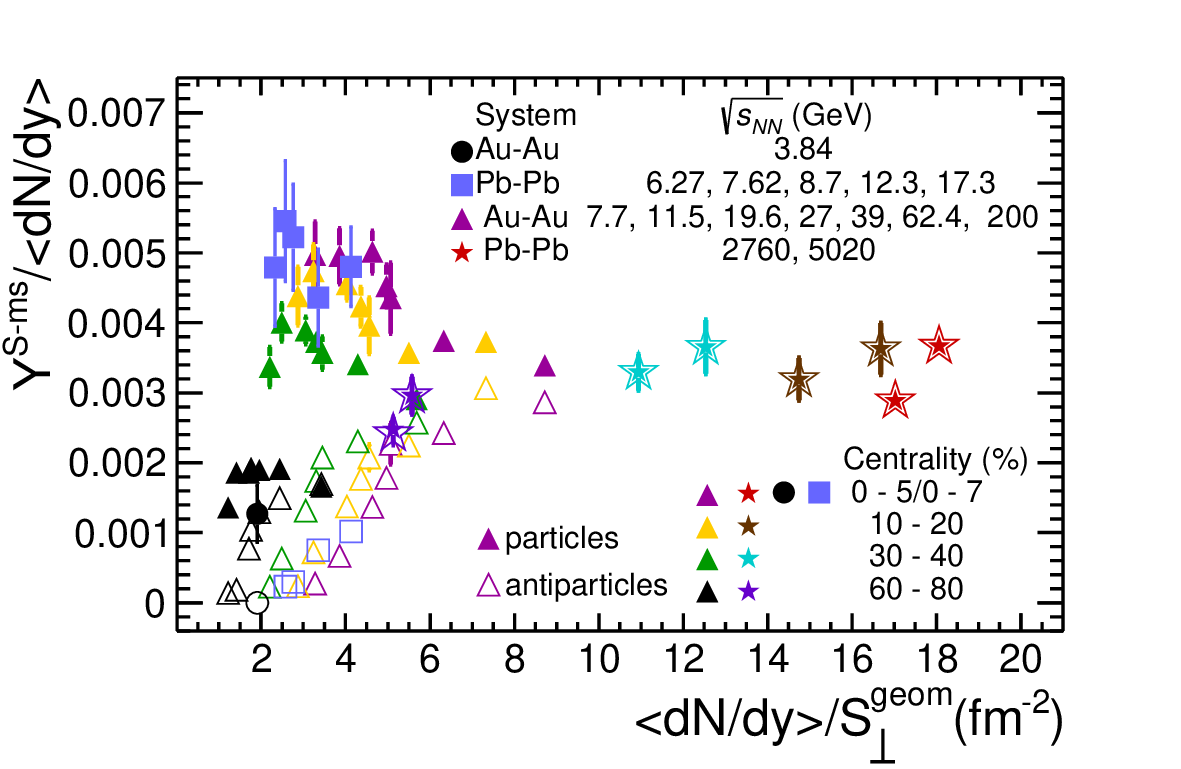}
\caption{The ratio of multi-strange hadron (full symbols)
and multi-strange anti-hadron (open symbols) yields per unit rapidity to the total particle multiplicity per unit rapidity as a function of entropy density
for four values of centrality.
}
\label{fig-19}
\end{figure}  

The data for strange hadrons exist at AGS energies only for central collisions. Nevertheless, a maximum in $Y^{S-1s}/\langle dN/dy \rangle$ and $Y^{S-ms}/\langle dN/dy \rangle$ in the case of particles as a function of entropy density starts to be evidenced going from central to peripheral collisions. While the entropy density corresponding to the maximum decreases from central to peripheral collisions, the corresponding value of the collision energy is slightly increasing, as can be seen in Figure \ref{fig-17}. Such a behaviour is in a qualitative agreement with the results of the parton-hadron string dynamics (PHSD)  transport model \cite{bratkovskaya} which predicts a slight increase in the collision energy at which the maximum in the $K^{ +}/\pi^{+}$ ratio versus the collision energy representation appears going from Au-Au to C-C central collisions. However, one should mention that the model does not evidence a similar trend in the $(\Lambda + \Sigma^{0})/\pi$ ratio as a function of collision energy for the three symmetric colliding systems, i.e. Au-Au, Ca-Ca and C-C. The decrease of the ratio of the strange hadron yield per unit rapidity to the total particle multiplicity per unit rapidity for peripheral collisions supports the conclusion of the same approach that the main contribution for the observed increase at lower entropy density/collision energy is due to chiral symmetry restoration which should decrease towards peripheral collisions, where the transparency is larger and the baryon density of the fireball is lower.
Focusing only on the more complete data for central collisions a kink is clearly evidenced in the correlation for single- and multi- strange hadrons around
the entropy density of $\approx$ 2.5 - 3 $fm^{-2}$. Beyond this value, 
$Y^{S}/\langle dN/dy \rangle$
decreases and converges to the value
of single- and multi-strange anti-hadrons at $\sqrt{s_{NN}}$ = 200 GeV. 
For strange anti-hadrons, a steep increase at lower collision energies is followed by a levelling off and another increase is observed at the largest LHC energy, more pronounced in the case of multi-strange anti-hadrons.
The same type of representation for kaons can be followed in Figure \ref{fig-20}. Apart the absolute values, the trends in the $\langle dN/dy \rangle/S_{\perp}$ dependence of the ratios of the $K^{+}$, $K^{-}$ yields per unit rapidity to the total particle multiplicity per unit rapidity  are similar with those corresponding to single- and multi-strange hadrons and anti-hadrons represented in Figures \ref{fig-18} and \ref{fig-19}, the kink location in $\langle dN/dy \rangle/S_{\perp}$ being the same. $K^{0}_{S}$ follows more or less the same trend as $K^{-}$ as its formation requires the presence of strange quark-antiquark pairs. The same type of representation for the $\Lambda$ and $\bar{\Lambda}$ ratios of yields per unit rapidity to the total particle multiplicity per unit rapidity as a function of entropy density, presented in Figure \ref{fig-21}, shows that the kink position is the same as in the case of Figures \ref{fig-18}, \ref{fig-19} and \ref{fig-20}.
Figure \ref{fig-22} is the same type of representation for the ratio of the $\Xi^{-}$, $\bar{\Xi^{+}}$ yields per unit rapidity to the total particle multiplicity per unit rapidity as a function of entropy density. 
\begin{figure} [htbp]
\includegraphics[width=1.\linewidth]{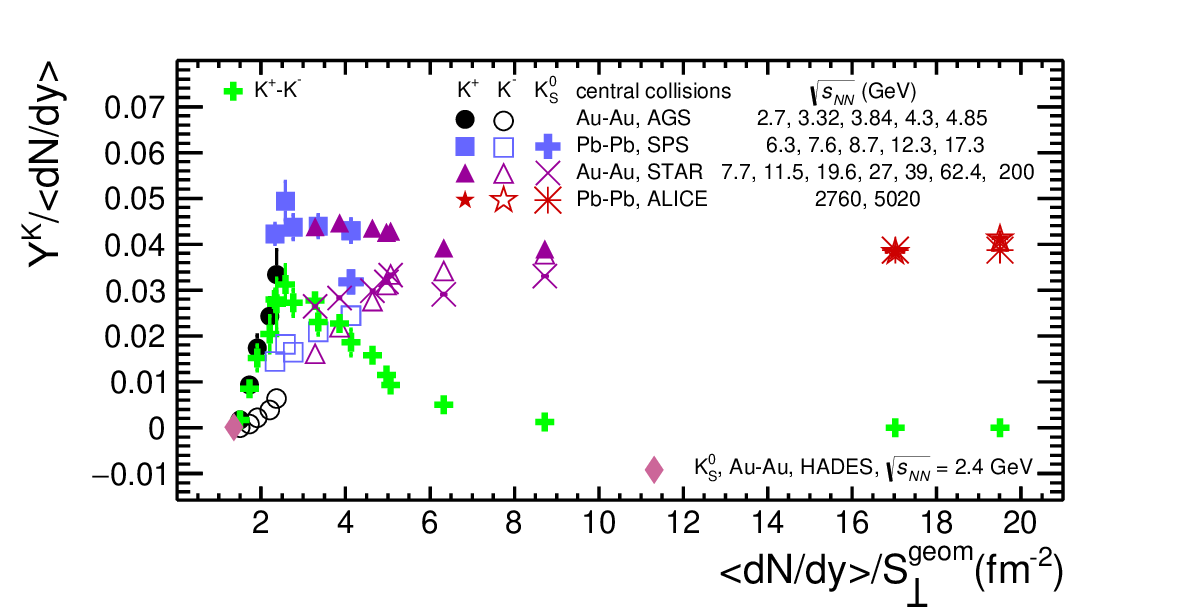}
\caption{The ratio of the yields per unit rapidity to the total particle multiplicity per unit rapidity as a function of entropy density  for $K^{+}$ (full symbols), $K^{-}$ (open symbols),
their difference (green crosses) and $K^{0}_{S}$, from AGS to LHC energies, for central collisions. 
}
\label{fig-20}
\end{figure} 
\begin{figure} [htbp]
\includegraphics[width=1.\linewidth]{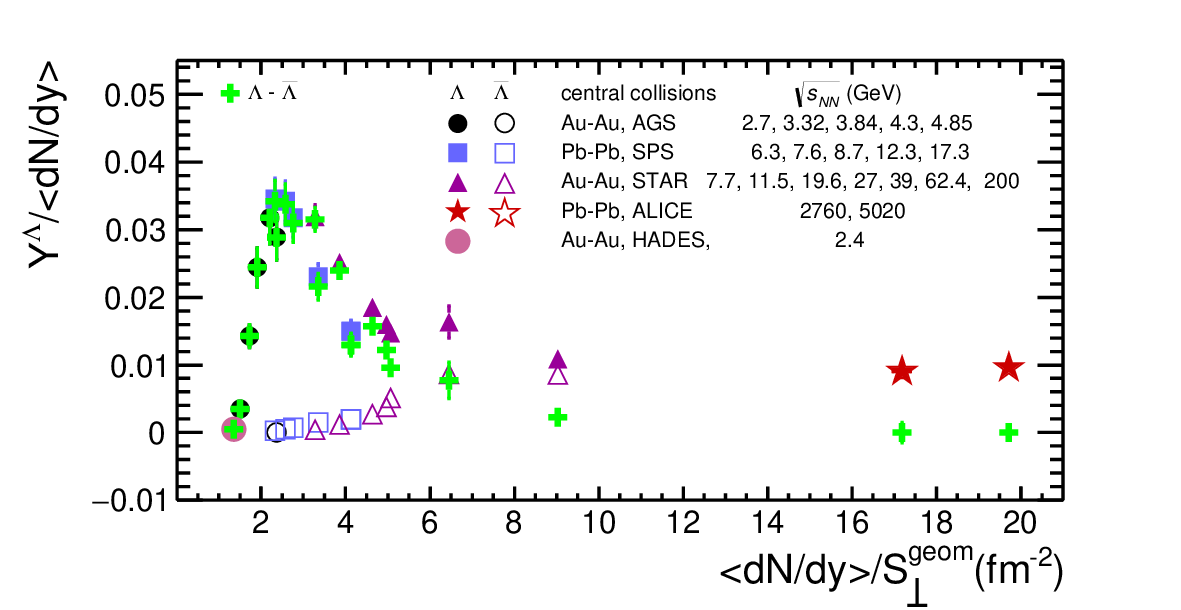}
\caption{The ratio of the yields per unit rapidity to the total particle multiplicity per unit rapidity as a function of entropy density  for $\Lambda$ (full symbols), $\bar{\Lambda}$ (open symbols),
their difference (green crosses) from AGS to LHC energies, for central collisions. 
}
\label{fig-21}
\end{figure}  
\begin{figure} [htbp]
\includegraphics[width=1.\linewidth]{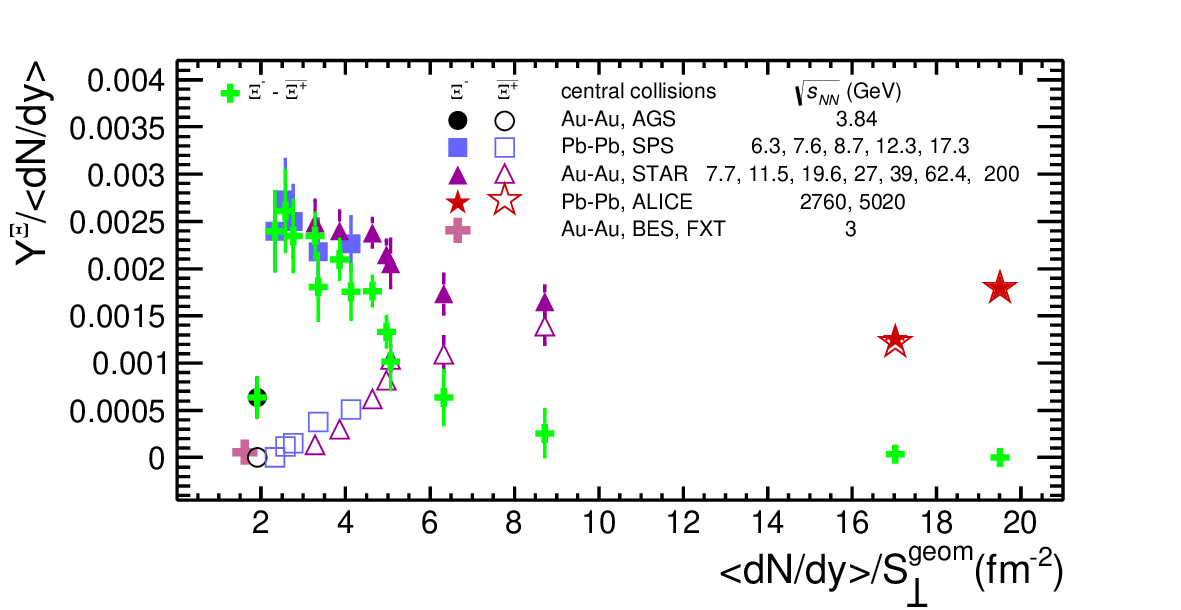}
\caption{The ratio of the yields per unit rapidity to the total particle multiplicity per unit rapidity as a function of entropy density  for $\Xi^{-}$ (full symbols), $\bar{\Xi^{+}}$ (open symbols), their difference (green crosses) from AGS to LHC energies, for central collisions. 
}
\label{fig-22}
\end{figure}  
The difference between the ratios of the yield per unit rapidity to the total particle multiplicity per unit rapidity corresponding to hadrons and anti-hadrons is represented by the green crosses in Figures \ref{fig-20} - \ref{fig-22}.
In such representation the kink is better evidenced, its location in $\langle dN/dy \rangle/S_{\perp}$ remaining unchanged at around
the entropy density of $\approx$ 2.5 - 3 $fm^{-2}$. 
All these representations also highlight the fact that the new data from HADES \cite{HADES} and BES fixed-target (FXT) \cite{BESFXT} experiments fit well into the general systematics. 
Such a sharp peak, evidenced for the first time at SPS based on data from AGS and SPS, in the $K^{+}/\pi^{+}$ ratio as a function of $\sqrt{s_{NN}}$ for central collisions, coined "horn" was considered to be a signature of a phase transition from a hadron gas to QGP \cite{gazd1,alt}. Within the framework of a thermal model \cite{Cleym}  
it was shown that such a kink type behaviour is related to a transition from a baryon-dominated hadronic gas to a meson-dominated one. Based on the PHSD approach  \cite{Cassing} the enhancement in the $K^{+}/\pi^{+}$ and $(\Lambda+\Sigma^{0})/\pi^{-}$ ratios evidenced at $\sqrt{s_{NN}}$ $\approx$ 6 GeV is caused by chiral symmetry restoration while deconfinement is essential in producing the maximum. 
\begin{figure} [htbp]
\includegraphics[width=1.\linewidth]{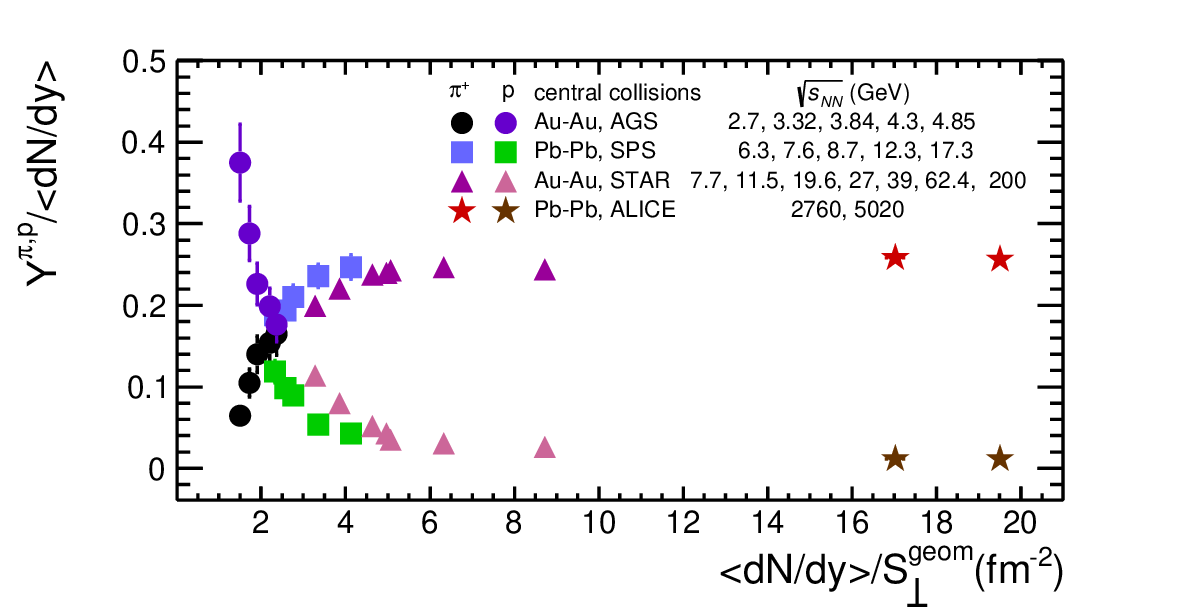}
\caption{Entropy density dependence of the ratio of the yield per unit rapidity  to the total particle multiplicity per unit rapidity of pions and protons for central collisions.  
}
\label{fig-23}
\end{figure} 
In Figure \ref{fig-23} is represented the dependence on the entropy density of the ratio of the yield per unit rapidity  to the total particle  multiplicity per unit rapidity for pions and protons for central collisions. The ratio of the proton yield per unit rapidity to the total particle multiplicity per unit rapidity decreases up to the top RHIC energy, becoming negligible  at the LHC energies. The ratio of the pion yield per unit rapidity to the total particle multiplicity per unit rapidity increases up to the top RHIC energy and  remains constant up to the LHC energies. They cross each other at the entropy density of $\approx$ 2.5 - 3 $fm^{-2}$ where a transition from a baryon-dominated  to a meson-dominated matter takes place \cite{Cleym}. This value is in the region corresponding to the position of the kinks evidenced in Figures \ref{fig-18} - \ref{fig-22} and corresponds to  $\sqrt{s_{NN}}$ $\approx$ 6 GeV. It is well below the energy where a plateau is reached in the $\langle dE_{T}/dy \rangle/\langle dN/dy \rangle - \langle dN/dy \rangle/S_{\perp}$ correlation or in the collision energy dependence of the slope of the Bjorken energy density times the interaction time as a function of $\langle dN/dy \rangle/S_{\perp}$ \cite{petro1}, specific to a transition from a hadronic phase to a coexistence of hadronic and deconfined matter. In view of the results obtained using the PHSD transport approach \cite{Cassing}, the strangeness enhancement up to the maximum is due to chiral symmetry restoration at high baryon density. Therefore, the position of the maximum could be indicative for a maximum of the baryon density of the fireball \cite{Cleym,busza}. Contrary to the expectations based on the statistical model \cite{Cleym} the kink position stays at the same value of entropy density/collision energy for $K$, $\Lambda$ and $\Xi^{-}$, no mass dependence being observed. The centrality dependence of the kink position evidenced in the $Y^{S}/\langle dN/dy \rangle$ - $\langle dN/dy \rangle/S_{\perp}$ correlation for particles also excludes its interpretation as signature for critical point \cite{stock}.
\section{Comparison with the results from $pp$ collisions at LHC energies} 
Measurements in pp collisions up to very high charged particle multiplicities at LHC energies, revealed
similarities between pp and Pb-Pb in many respects, i.e.
near-side long range pseudorapidity correlations \cite{CMS1}, the
correlation among the parameters extracted from  the
$p_{T}$ distributions fitted with a BGBW formula 
($\langle \beta_{T} \rangle - T ^{fo}_{kin}$) as a function of charged particle
multiplicity \cite{Cristi1}, azimuthal angular correlations \cite{CMS2}, geometrical scaling \cite{petro3,petro4,Pet2}, etc. In order to have 
an unambiguous conclusion on the origin of such similarities, the studies of the types of correlations presented in the previous chapters for A-A collisions are extended to pp. 
Experimental data on transverse momentum spectra for light flavors as a function of the average charged particle 
multiplicity at LHC energies were published in \cite{ALICE9,ALICE10,ALICE11}.
\begin{figure} [htbp]
\includegraphics[width=1.\linewidth]{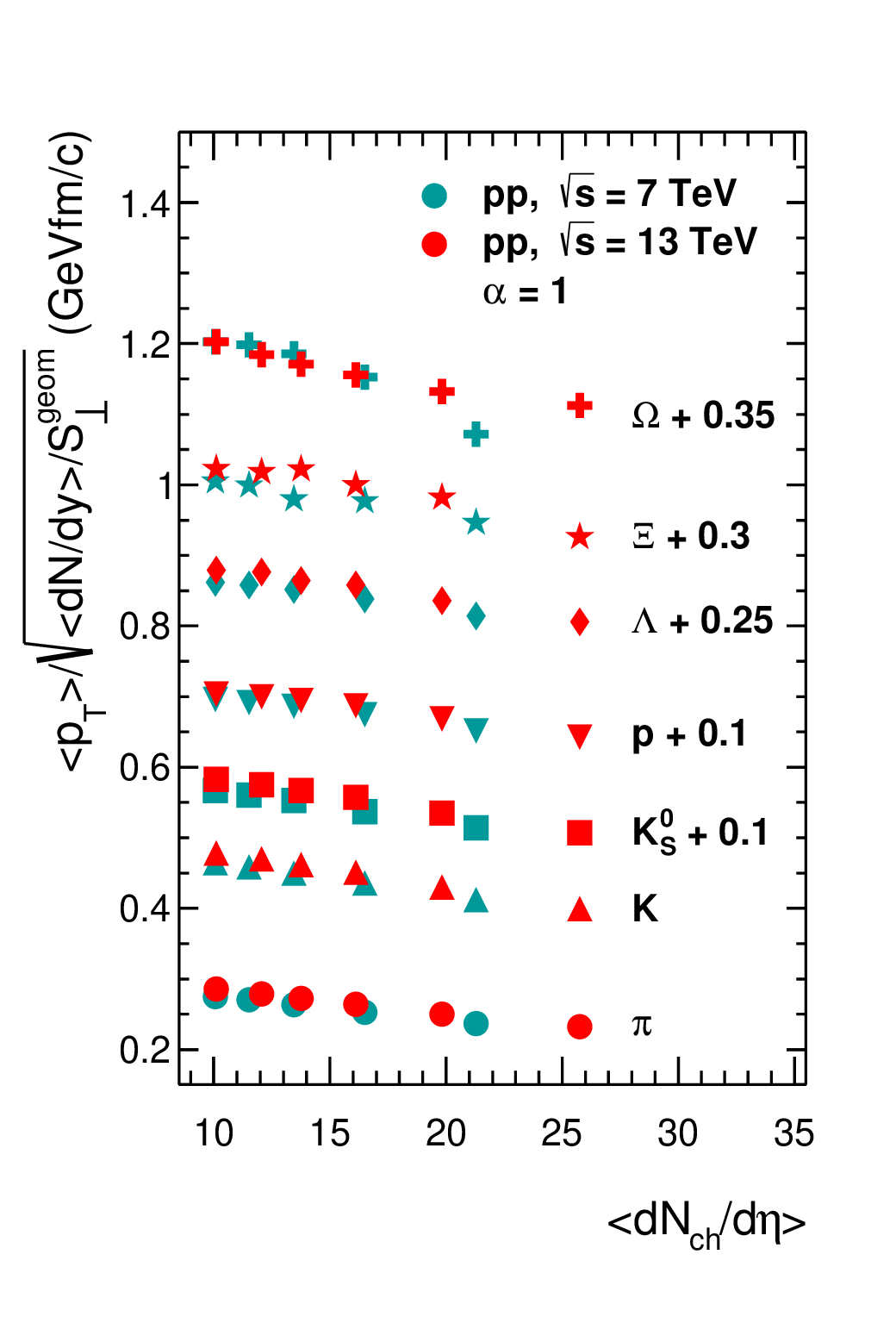}
\caption{The $\langle dN_{ch}/d\eta \rangle$  dependence of $\langle p_{T} \rangle/\sqrt{\langle dN/dy \rangle/S_{\perp}}$  for $\pi$, K, $K^{0}_{S}$, p, $\Lambda$, $\Xi$ and $\Omega$  measured in  pp collisions at $\sqrt{s}$ =7 TeV and $\sqrt{s}$ = 13 TeV.
}
\label{fig-24}
\end{figure} 
In order to calculate the entropy density, an estimate of the transverse overlap area in the case of a pp collision as a function of charged particle multiplicity per unit pseudorapidity is required. 
As in the previous papers \cite{petro3,petro4,Pet2}, 
the transverse overlap area for pp collisions $S_{\perp}^{pp}$=$\pi$
$r_{max}^2$ was
estimated using the result of \cite{Bzdak}, computed in the IP-Glasma 
model. In this approach
$r_{max}$ is the maximal radius for which the energy density of the 
Yang-Mills fields is above 
$\varepsilon=\alpha\Lambda_{QCD}^4$ ($\alpha\in{[1,10]}$).
The model results were fitted for $\alpha$ =1, with: 
\begin{equation}\label{eq4}
f_{pp}=
\left\{
\begin{array}{rl}
0.387+0.0335x+0.274x^2-0.0542x^3 & \mbox{if $x<3.4$}\\
1.538                            & \mbox{if $x\geq$ 3.4} 
\end{array}
\right.
\end{equation}
where, by approximating the gluon density with the total particle multiplicity per unit rapidity x=$(dN/dy)^{1/3}$, and $r_{max}$=1fm$f_{pp}(x)$. 
In Figure \ref{fig-24} is presented the dependence of
$\langle pT \rangle/\sqrt{\langle dN/dy \rangle/S_{\perp}}$ on the charged particle multiplicity per unit pseudorapidity ($\langle dN_{ch}/d\eta \rangle$) for $\pi$, K, $K^{0}_{S}$, p, $\bar{\Lambda}$, $\Xi$ and $\Omega$  for pp collisions at $\sqrt{s}$ =7 TeV and $\sqrt{s}$ = 13 TeV. A clear decrease as a function of $\langle dN_{ch}/d\eta \rangle$ is observed. The transverse overlap area was considered to be the same for all species, corresponding to $\alpha$ = 1 ( Eq. \ref{eq4}). Qualitatively the trend is similar with the one corresponding to A-A collisions (see Figure \ref{fig-1} of the present paper and Figure 1 of Ref. \cite{petro1}), supporting the model predictions \cite{levin,lappi}. 
\begin{figure} [htbp]
\includegraphics[width=1.\linewidth]{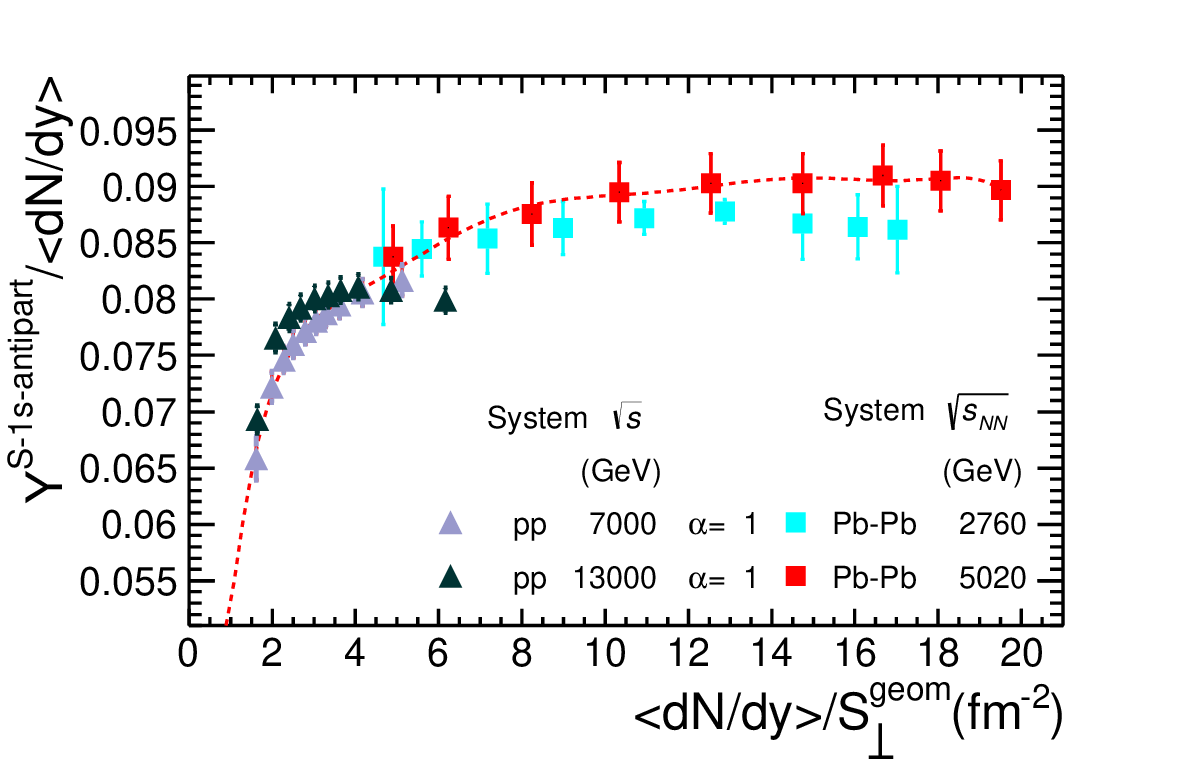}
\caption{The ratio of the yield of single-strange anti-hadrons to the total particle multiplicity per unit rapidity as a 
function of entropy density for Pb-Pb and pp collisions at the LHC energies (with dashed line as a guide to the eye). Up-triangles correspond to   
 $\alpha$ = 1 for estimating the transverse overlap area in pp collisions (see text). 
}
\label{fig-25}
\end{figure}  
\begin{figure} [htbp]
\includegraphics[width=1.\linewidth]{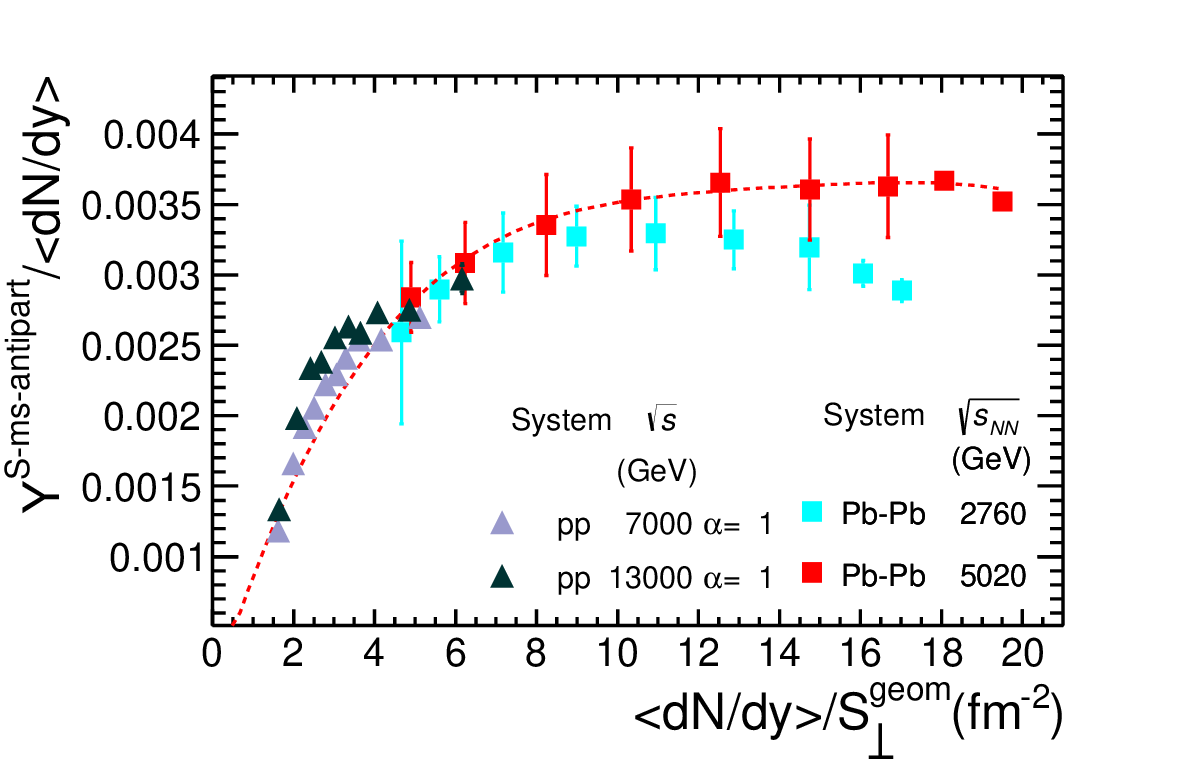}
\caption{The ratio of the yield of multi-strange anti-hadrons to the total particle multiplicity per unit rapidity as a 
function of entropy density for Pb-Pb and pp collisions at the LHC energies (with dashed line as a guide to the eye). Up-triangles correspond to   
 $\alpha$ = 1 for estimating the transverse overlap area in pp collisions (see text). 
 }
\label{fig-26}
\end{figure}  
As far as in the case of pp collisions at LHC energies the production of pairs of quarks and antiquarks is the main process, the difference in the representation of the $Y^{S}/\langle dN/dy \rangle$ - $\langle dN/dy \rangle/S_{\perp}$ correlations between strange hadrons and strange anti-hadrons is minor.
Plotting the results for $Y^{S-1s}/ \langle dN/dy \rangle$ and  $Y^{S-ms}/ \langle dN/dy \rangle$ in the case of anti-hadrons for pp collisions at $\sqrt{s}$ = 7 TeV and  $\sqrt{s}$ = 13 TeV along with the corresponding data for Pb-Pb at LHC energies,
Figures \ref{fig-25} and \ref{fig-26}  are obtained. 
 The results corresponding to $\alpha$ = 1 are represented by up triangles. As can be seen in Figures \ref{fig-25} and \ref{fig-26},  
 for both single- and multi-strange anti-hadrons the pp data using $\alpha$ = 1 for estimating $S_{\perp}$ are laying on top of the extrapolation of Pb-Pb results towards smaller values of the entropy density, the dashed lines being only a guide to the eye.
\section{Conclusions}
The present paper is focused on  strange hadron production in heavy ion collisions and similarities between pp and Pb-Pb at LHC energies. This study is based on the impressive amount of experimental information obtained in A-A collisions starting from the AGS energies up to the LHC ones. For strange hadrons, the observed trends in the ratio between the average transverse momentum and the square root of the total particle multiplicity per unit rapidity and unit transverse overlap area $\langle p_{T} \rangle/\sqrt{\langle dN/dy \rangle/S_{\perp}}$  as a function of collision energy for a given centrality or as a function of centrality for a given collision energy are similar with the results for $\pi$ , K and p presented in a previous study, supporting the predictions of CGC and percolation based approaches. The slope and offset of the $\langle p_{T} \rangle$ dependence on the particle mass and the parameters of BGBW simultaneous fits of the $p_{T}$ spectra for strange and multi-strange hadrons, the average transverse expansion velocity ($\langle \beta_{T} \rangle$) and the kinetic freeze-out temperature ($T_{kin}$), are extracted and compared to the results of the same analysis for $\pi$ , K and p. Represented as a function of $\sqrt{\langle dN/dy \rangle/S_{\perp}}$, the larger slope and  $\langle \beta_{T} \rangle$ values  for $\pi$, K and p relative to the ones corresponding to strange- and multi-strange hadrons and the smaller offset  and $T_{kin}$ values for $\pi$, K and p relative to those corresponding to strange and multi-strange hadrons support the dynamical freeze out scenario. $K^{0}_{S}$, $\Lambda$, $\Xi^{-}$ and $\Omega^{-}$ suffer negligible interaction with the hadron environment conserving the momentum distribution corresponding to the their formation time, while the $\pi^{+}$ , $K^{+}$ and p build up extra expansion via elastic interaction with the hadronic medium which cools down due to expansion. 

The detailed study of the entropy density dependence of the ratio of strange hadron yields per unit rapidity to the total particle multiplicity per unit rapidity at different collision energies and centralities reveals the necessity to study separately  single-, multi- strange hadrons and anti-hadrons, in the multi-strange case the sensitivity to the composition of the fireball being higher. 
The ratio of the strange anti-hadron yields per unit rapidity  to the total particle multiplicity per unit rapidity 
as a function of entropy density studied as a function of the fireball size shows a similarity with the correlation between the ratio of the energy density to the entropy density as a function  of entropy density \cite{petro1}. A kink type behaviour in the ratio of the single-,  multi- strange hadrons, $K^{+}$, $\Lambda$ and $\Xi^{-}$  yields per unit rapidity to the total particle multiplicity per unit rapidity and in the difference between particles and anti-particles, showing up in the region where a transition from a baryon-dominated gas to a meson-dominated one takes place, is evidenced, in agreement with the expectations based on a thermal model \cite{Cleym}. In this energy range the baryon transparency sets-in and a hot mesonic fireball starts to be formed by particle production in the collision as precursor of deconfinement. 
The maximum in $Y^{S-1s-part}/\langle dN/dy \rangle$ and $Y^{S-ms-part}/\langle dN/dy \rangle$ as a function of entropy density starts to be evidenced going from central to peripheral collisions, where only the data from RHIC and LHC were taken into account. While the entropy density corresponding to the maximum decreases from central to
peripheral collisions, the corresponding value of the collision energy is slightly increasing.
Within the experimental error bars, the position of this maximum does not depend on the mass of the corresponding strange hadrons. Similar with A-A collisions, a clear decrease of $\langle p_{T} \rangle/\sqrt{\langle dN/dy \rangle/S_{\perp}}$ with $\langle dN_{ch}/d\eta \rangle$ for $\pi$, K, $K^{0}_{S}$, p, $\Lambda$, $\Xi$  and $\Omega$  for pp collisions at $\sqrt{s}$ = 7 TeV and $\sqrt{s}$ = 13 TeV is evidenced supporting the expectations based on CGC and percolation models. In the comparison between Pb-Pb and pp for single- and multi- strange anti-hadrons, the pp results nicely follow the extrapolation of Pb-Pb data at lower entropy density. \section{Acknowledgements}
The work was carried out under the contracts sponsored by the Ministry of Research, Innovation and Digitization: RONIPALICE-07/03.01.2022 (via IFA Coordinating Agency) and PN-23 21 01 03.
%\newpage
\nocite{*}
\bibliography{draft_strangeness}% Produces the bibliography via BibTeX.
\end{document}